**Piles of piles: An inter-country comparison of nuclear pile development during World War II**

B. Cameron Reed
Department of Physics (Emeritus)
Alma College
Alma, Michigan 48801 USA
reed@alma.edu

January 23, 2020

**Abstract**

Between the time of the discovery of nuclear fission in early 1939 and the end of 1946, approximately 90 "nuclear piles" were constructed in six countries. These devices ranged from simple graphite columns containing neutron sources but no uranium to others as complex as the water-cooled 250-megawatt plutonium production reactors built at Hanford, Washington. This paper summarizes and compares the properties of these piles.

1.  **Introduction**

According to the World Nuclear Association, there were 448 operable civilian nuclear power reactors in the world with a further 53 under construction as of late 2019.[1] To this total can be added reactors intended for other purposes such as materials testing, medical isotope production, operator training, naval propulsion, and fissile materials production.

All of these reactors are the descendants, by some path or other, of the first generation of nuclear "piles" developed in the years following the discovery of nuclear fission in late 1938. Any historian of science possessing even only a passing knowledge of developments in nuclear physics during World War II will be familiar with how Enrico Fermi achieved the first self-sustaining chain reaction with his CP-1 ("Critical Pile 1") graphite pile at the University of Chicago on December 2, 1942, and how this achievement led to the development, within two years, of large-scale plutonium production reactors at Hanford, Washington. Those familiar with more of the story of reactor development during World War II will further know that German scientists were also conducting their own pile experiments, but that by the end of the war they had fallen far behind their Allied counterparts, not having achieved even a low-power self-sustaining reaction.

The great success of the Manhattan Project has tended to make accounts of wartime developments in nuclear physics seem a very one-sided affair. But as often happens, the real circumstances were more complex. When an acquaintance asked me how many piles had been constructed by the end of the war, my initial response was along the lines of "Perhaps a few dozen pre-CP-1 experimental models constructed by the Fermi group, plus a few on the German side." But the query piqued my curiosity, which led to a literature search and a surprising answer: Between 1939 and the end of 1946, some 90 piles had been developed in six countries. The vast majority, about 60, were indeed located in the United States, but some 20 were constructed in Germany, with a few others in France, Britain, Canada, and Russia. I took the end date of my search to be the end of 1946 in order to include the first Russian pile.

While the individual-country stories of these piles are accessible to various degrees, they have not to my knowledge been gathered in one place. This is the purpose of this paper: To summarize the characteristics and purposes of the ~ 90 piles, and to provide readers who may wish to explore the details of particular ones in more depth with a list of appropriate references.

A number of caveats need to be stated here. First, this survey is surely incomplete. It is known, for example, that Samuel Allison had initiated pile experiments at the University of Chicago before Enrico Fermi arrived there, but I have not been able to find any detailed descriptions of his work. Second, enumerating exactly how many piles were developed depends on how one counts. Should a pile which contained no uranium but was used for the important job of testing neutron diffusion through different types of graphite be counted? Should a pile which was dismantled and then re-built, say with a different configuration of uranium or graphite, be counted as one or multiple contributions to the total? My figure of ~ 90 is based on answering both of these questions in the affirmative on the rationale that the work demanded real planning and physical effort on the part of the investigators involved and that each pile played a role in informing future research. Third, the level of information available for individual piles can vary widely. Dates of construction can sometimes only be estimated to on the order of a season or a month; dimensions and amounts of materials were sometimes reported in metric units, sometimes in customary American units, and sometimes not at all or had to be inferred; "uranium oxide" could mean $UO_2$ or $U_3O_8$; and some piles could be assigned an identification only by the number of a report in which they are discussed. I have tried as much as possible to retain the units reported in original documents and to make clear what quantities have been estimated after-the-fact.

With so many piles involved, there is also the question of how to organize the information. A chronological ordering would be attractive from the point of view of comparing developments at different locations, but this would be extremely confusing as so much parallel development went on in the United States and Germany (see the timeline in Fig. 1). I decided on a country-by-country approach of three tables: One for France, Britain, Canada and Russia (Table I), one for the United States (Table II), and one for Germany (Table III). I have available a large spreadsheet comprising all accumulated information, which I would be happy to share with interested readers.

The structure of this paper is as follows. In Section 2 I summarize the sources of information on each country's piles, give a brief discussion of the process of nuclear fission and how nuclear reactors differ from nuclear weapons, and also a present a summary of the fundamental organization structures of the German and American nuclear programs. As the history of fission and the administrative aspects of these programs are well-covered in other sources, these summaries are intended to serve as brief introductions for readers who are unfamiliar with these aspects of this history; readers familiar with these issue will likely want to skip over these sections. Section 3 forms the heart of this paper: Essential data are summarized in Tables I-III, and details are elaborated in the text. To keep the Table for the United States piles to a reasonable length I have grouped some piles according to the original report wherein they are summarized. Section 4 offers a few closing remarks.

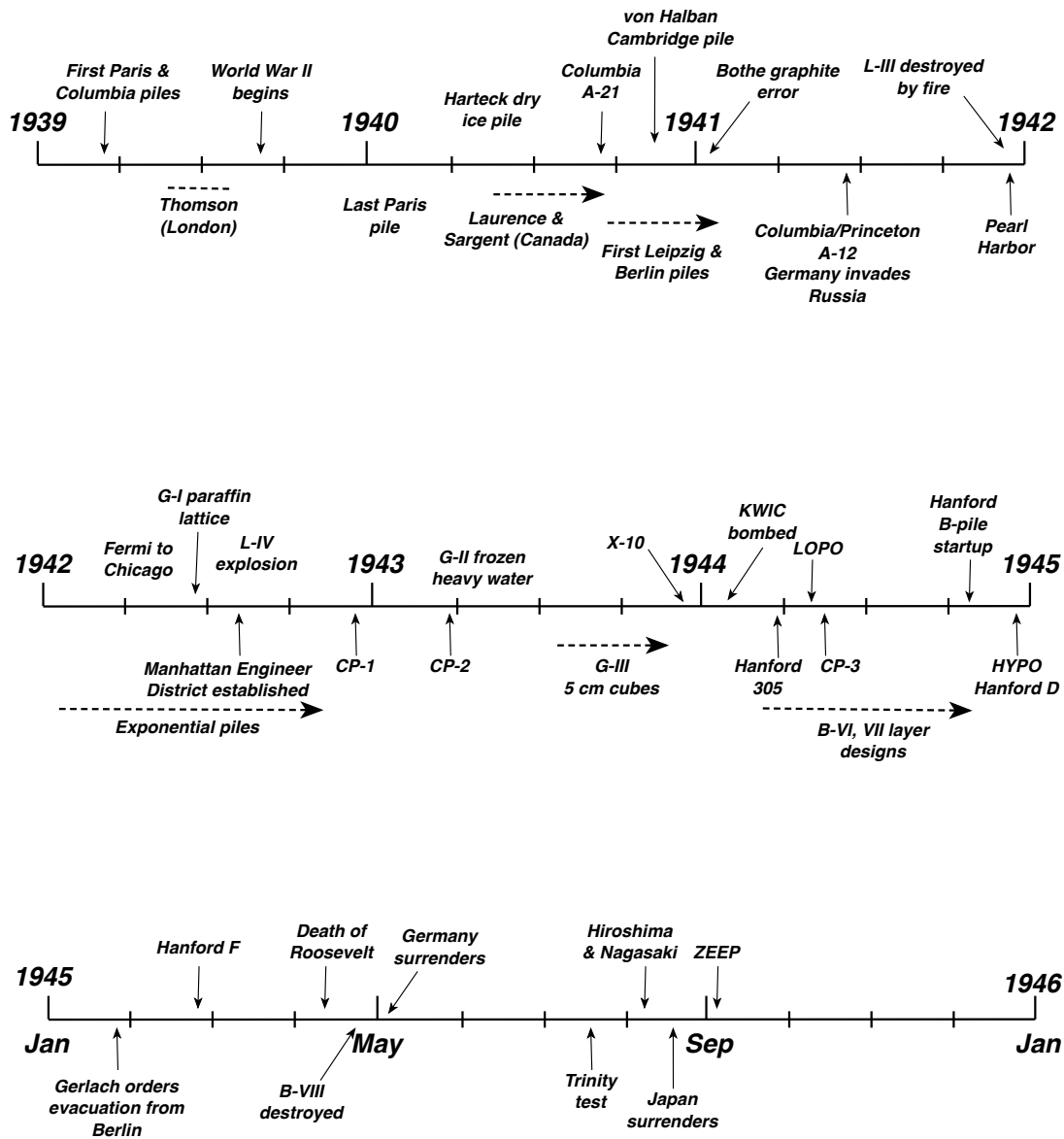

**Fig. 1.** Timeline of significant pile developments, 1939-1945. Not all piles listed in tables I-III are indicated here; some dates are approximate.

## 2. Sources, Fission, and Funding

### 2.1 Sources of information on pile developments

Information on early pile experiments is distributed among disparate sources. Some of the first pile experiment were performed in France by Hans von Halban and his collaborators, with some results appearing in publications such as *Nature* and *Journal de Physique et Radium*, but the most extensive English-language summary of these efforts appears in Spencer Weart's book on the development of nuclear energy in France, *Scientists in Power*.[2] Some experiments were carried out at about the same time by George P. Thomson in Britain; these are described in Ronald Clark and Margaret Gowing's books on the British nuclear program.[3, 4] The very distinguished contributions to reactor development in Canada are ably related in Wilfrid Eggleston's *Canada's Nuclear Story* and a publication of Atomic Energy of Canada Limited.[5, 6]

Much primary literature is available on the American program. Before wartime censorship went into effect, some papers were published in the *Physical Review*, but the essential source of information here is Volume II of the collected works of Enrico Fermi.[7] This volume covers the time from Fermi's arrival in the United States in early 1939 to his death in 1954, and includes copies of his reports to various administrative bodies such as the Uranium Committee and the National Defense Research Committee (NDRC), which in June 1941 became the Office of Scientific Research and Development. Supplementing this is a paper published by Fermi in the December 1952 edition of *American Journal of Physics* in recognition of the tenth anniversary of the successful startup of CP-1, and material in David Schwartz's recent biography of Fermi.[8, 9]

The German nuclear program has been extensively analyzed by David Irving, Mark Walker, and Per Dahl; their work has been a boon to nuclear history scholars who lack convenient access to archives in that country.[10, 11, 12] Dahl also includes a brief chapter on the Canadian project.

For Western authors the most difficult history to penetrate is always that of Russian/Soviet developments. For the present purposes there is only one Russian pile of interest, the so-called F-1 pile, the design of which was based on obtaining the design of a Hanford pile through espionage. For this case I have relied on on-line sources and Richard Rhodes' description of this tale.[13, 14]

### 2.2 Fission, reactors, and bombs: A brief tutorial

Nuclear fission is the name given to the phenomenon that nuclei of some isotopes of selected heavy elements, most notably those of the uranium isotope of atomic weight 235 ($^{235}$U) can break apart when struck by a bombarding neutron. In this process, which was discovered in Berlin in late 1938, the uranium nucleus splits into two fragments, with the sum of the masses of the fragments being less than that of the struck nucleus. The lost mass corresponds to an enormous amount of energy via Albert Einstein's famous $E = mc^2$ equation. The amount of energy released per fission reaction is *millions* of times that liberated in any known chemical reaction; this is what makes nuclear explosives so compelling in comparison to conventional ones. Only a few weeks after the discovery of fission it was verified that a by-product of each such reaction was the liberation of two or three neutrons. It is this feature that makes a chain reaction possible: These "secondary" neutrons, if they do not escape the mass of uranium, can go on to fission other nuclei. Once this process is started it can in principle continue until all of the uranium is fissioned. Chain reactions are the fundamental process by which nuclear reactors and bombs function.

The discovery of fission opened numerous questions. Could any other elements undergo fission? Would some minimum amount of uranium have to be arranged in one place to achieve a chain reaction? If a chain reaction could be initiated, could it be controlled to provide power? By the time of the outbreak of World War II in September, 1939, understanding of the physics of fission was beginning to come into focus. For practical purposes (at least at the time), only uranium appeared to be fissile. However, the prospects for realizing nuclear energy on a practical scale looked dim. As it occurs naturally, uranium comprises two isotopes, U-235 and U-238. However, these isotopes occur in very unequal proportions: only about 0.7% of naturally-occurring uranium is of the U-235 variety, while the other 99.3% is U-238. By early 1940, experiments had established that only nuclei of the rare isotope U-235 had a useful likelihood of fissioning when bombarded by neutrons, whereas those of U-238 would tend to slow down and capture incoming neutrons without fissioning. Given the overwhelming preponderance of U-238 in nature, this capture effect promised to poison the prospect for a chain reaction. To obtain a chain reaction, it appeared that it would be necessary to isolate a sample of U-235 from its sister isotope, or process uranium so as to increase the percentage of U-235. Such process is now known as enrichment, and is what was used to isolate the U-235 used in the nuclear weapon dropped at Hiroshima. Enrichment is, however, an extremely difficult and expensive task, demanding a very large investment in precision-engineered infrastructure. Since isotopes of any element behave identically so far as their chemical properties are concerned, no chemical separation technique can be employed to achieve enrichment, only a method that depends on the very slight mass difference between the two isotopes could be a possibility. Enrichment on such a large scale had never been carried out, particularly with heavy elements, so in 1940 the prospect of a nuclear weapon any time soon looked more like an issue of science fiction than practical engineering.

The separate idea of a nuclear pile also began to originate soon after the discovery of fission, and began to grow in importance about the middle of 1940 due to the development of more refined understanding of the differing responses of the two uranium isotopes to bombarding neutrons. Scientists in Europe, Britain, and America realized that it might be possible to achieve a *controlled* (not explosive) chain reaction using natural uranium without enrichment as a consequence of the nuances of how nuclei react to bombarding neutrons. When a neutron strikes a nucleus, various reactions are possible: The nucleus might fission, it might capture the neutron and not fission, or it might simply scatter the neutron much as a billiard ball would deflect an incoming marble. Each process has some probability of occurring, and these probabilities can depend sensitively on the speed of the incoming neutrons. Neutrons released in fission reactions are extremely energetic, emerging with average speeds of about 20 million meters per second; these are so-called "fast" neutrons. As remarked above, U-238 tends to capture fast neutrons that have been emitted in fissions of U-235 nuclei. However, when a nucleus of U-238 is struck by a very slow neutron - one traveling at a mere couple thousand meters per second - it behaves as a much more benign target, with scattering preferred to capture by odds of about three-to-one. But for such slow neutrons, U-235 turns out to have an enormous fission probability: over 200 times greater than the capture probability for U-238. This factor is large enough to compensate for the small natural abundance of U-235, and is what renders a *slow neutron* chain reaction possible. The key concept here is that slowing fission-liberated neutrons is effectively equivalent to enriching the abundance percentage of U-235. Thus, if neutrons emitted in fissions can be slowed, they will have a good chance of going on to fission other U-235 nuclei before being lost to capture by U-238 nuclei. In actuality, both processes proceed simultaneously in an operating reactor. Counter-intuitively, neutron capture by U-238 nuclei turns out to be indirectly advantageous for bomb-makers, as is explained later in this section.

The trick to slowing a neutron during the very brief time between when it is born in a fission and when it strikes another nucleus is to work not with a single large lump of uranium but rather to disperse it as small chunks throughout a surrounding medium which slows neutrons without capturing them. Such a medium is known as a moderator, and the entire assemblage is now known as a reactor. During the war the synonymous term "pile" was used in the literal sense of a "heap" of metallic uranium slugs and moderating material. Ordinary water proved to capture too many neutrons to serve as a moderator, but both heavy water and graphite (crystallized carbon) proved suitable. By introducing moveable rods of neutron-capturing material into the pile and adjusting their positions as necessary, the reaction can be controlled to yield a source of energy. All modern power-producing reactors still operate via chain-reactions mediated by moderated neutrons. A reactor cannot be made into a bomb, however: The reaction is far too slow, and even if the control rods are rendered inoperative the reactor will melt itself long before blowing up: Fukushima.

The attractiveness of the pile concept during the was not as a weapon but as a source of propulsion. A nuclear reactor is unlike an internal combustion engine in that it does not require a constant supply of oxygen to operate. Propelled by a reactor, a submarine could remain submerged for much longer and have a much greater range than a conventionally-powered one. Nuclear-powered (and armed) vessels would not emerge until the 1950's, but the concept well-established beforehand. More crucial for the current story, however, it that another concept also emerged about mid-1940: That it might be possible to use a nuclear pile to synthesize an alternate nuclear explosive material that, once made, would not require any enrichment before being fashioned into a bomb.

It was mentioned above that neutron capture by U-238 nuclei will also proceed within an operating reactor. On capturing a neutron, a nucleus of U-238 becomes one of U-239. Based on extrapolating from experimentally-known patterns concerning the stability of nuclei, it was predicted that U-239 nuclei might decay within a short time to nuclei of atomic number 94, the element now known as plutonium, and that said element might be very similar in its fissility properties to U-235. If so, a reactor sustaining a controlled chain reaction via U-235 fissions could be used to "breed" plutonium from U-238. The plutonium could be separated from the mass of parent uranium fuel by conventional chemical means (they are different elements) and then used to construct a bomb; this is what obviates the need to develop enrichment facilities. Within months, these predictions were partially confirmed on a laboratory scale in the United States by creating a small sample of plutonium via moderated-neutron bombardment of uranium. Pile-generated plutonium would be used in the bomb dropped at Nagasaki.

## 2.3 Administrative Organizations

The contrast between the effectiveness of the German and American nuclear programs during the war speaks to the necessity that complex technical projects need not only capable scientists to carry out the work, but that they must also be effectively organized and managed. In this section I give brief histories of the organization of these two projects, beginning with the German one.

The German program got underway in April, 1939, as a result of a colloquium on fission given at the University of Göttingen by Wilhelm Hanle. Physicist Georg Joos of the same institution attended the colloquium and felt it his duty to inform government authorities of the possibilities for nuclear energy, and wrote to the Reich Ministry of Education, which oversaw universities. His letter reached Abraham Esau, a former academic physicist and Nazi supporter who was President of the Reich Bureau of Standards and head of the physics section of the Ministry's Reich Research Council. Esau organized a conference held in Berlin on April 29, of which one result was a recommendation that all uranium stocks in Germany be secured and banned from export.

Unknown to Esau, however, a second initiative was underway. On April 24, University of Hamburg physical chemist Paul Harteck and his assistant Wilhelm Groth had written a letter to the German War Office to alert them to the fact that developments in nuclear physics could lead to very powerful explosives. Their letter reached the Army Ordnance Department of the War Office, where it was routed to physicist Erich Schumann, an advisor to General Wilhelm Keitel, Chief of the Armed Forces High Command. Schumann delegated the issue to Kurt Diebner, an Army expert on nuclear physics and explosives. Thus, two rival programs were underway in Germany at the start of the war: Esau's and Diebner's. However, Esau would soon be sidelined by the much more powerful War Office bureaucracy: Within days of the start of the war in September he was informed that the Ordnance Department was ordering the Bureau of Standards to cease uranium research and that a cache of uranium oxide that Esau had accumulated would be appropriated.

The War Office's effort ramped up quickly. The same week as Esau's program was being poached, Erich Bagge, a physicist at the Leipzig Institute for Theoretical Physics and a student of Werner Heisenberg, was ordered to report to Army Ordnance in Berlin. There he was met by Diebner and Schumann, who wanted his help in arranging a conference of experimental physicists to explore the feasibility of using uranium as a source of power or explosives. This meeting, which was held on September 16, 1939, was attended by some of the leading experimental nuclear physicists in Germany, including Walther Bothe, Hans Geiger, Otto Hahn (a co-discoverer of fission), and Diebner. Esau was not on the guest list. Despite skepticism of achieving a chain reaction in view of the scarcity of uranium 235, two important developments came out of this meeting. First, Schumann recommended to General Karl Becker, head of the Army Ordnance Office, that a "Nuclear Physics Research Group" be established within the that department; this would lead to a research laboratory located in Gottow (a suburb of Berlin); Diebner was placed in charge of this initiative. Second, Bagge suggested that Heisenberg be brought in to work out the theory of chain reactions. A second conference ten days later saw things begin to move on a number of fronts. By this time Heisenberg had appreciated that two routes to utilizing fission might be possible: In reactors if a suitable neutron-moderating substance could be found, and/or as an explosive if U-235 could be isolated. Diebner and Bagge drew up a research program: Heisenberg would continue theoretical investigations of chain reactions, Bagge would undertake measurements of the neutron-collision properties of heavy-water (for use as a possible moderator), and Harteck would continue with isotope-separation work. Also at about this time, Schumann moved to have the War Office take over the facilities of the Kaiser-Wilhelm Institute of Physics (KWIP) in Berlin as a location at which to centralize the work. The Institute could hardly refuse, and its Dutch-born director, Peter Debye, left for America in January, 1940. Institute staff wanted Heisenberg to be appointed director, but Schumann designated Diebner instead, with Heisenberg to serve as an advisor. This move resulted in two pile-experiment groups being operative: Heisenberg and his collaborators at Leipzing, and Diebner's group at Gottow.

During 1941 the German program began to experience serious setbacks. In January, Walther Bothe concluded that graphite captured too many neutrons to make it suitable for use as a moderator, an experimental error which caused the German program to turn to hard-to-obtain heavy water as a moderator. The situation worsened when, by October of that year, German forces were becoming bogged down in the battle of Moscow. The economy was becoming strained, and, just two days before Pearl Harbor, Schumann wrote to the directors of institutes engaged in uranium research to tell them that work on the project could only be justified "if a certainty exists of attaining an application in the foreseeable future." A conference on December 16 resulted in a decision by the Army to seriously reduce its funding and relinquish control of the project back to the Reich Research Council. Abraham Esau thus came back into control of the initiative he had tried to stimulate almost three years earlier, but the Army team under Diebner continued its research at Gottow.

June 6, 1942 was a pivotal day for the fate of the German nuclear program. That day, Heisenberg met with Minister of Munitions Albert Speer and his staff to decide on the future of nuclear research; also present were Diebner, Harteck, Otto Hahn, Chief of Army Ordnance General Emil Leeb, and Field Marshall Erhard Milch, who was in charge of German aircraft production. Heisenberg addressed the group on possible military aspects of fission, but advised that it would be impossible for Germany to produce a bomb as no method enriching uranium on a large scale was available. Speer limited his approvals to various construction projects, including a bunker on the grounds of the KWIP which would be equipped to house a large pile. In further reorganizing, the Reich Research Council was placed under the directorship of Hermann Göring, who appointed Rudolf Mentzel, a civil servant, to manage the Council's affairs. Mentzel in turn delegated administration to various directors and "plenipotentiaries" for important research projects, with nuclear physics being placed under the purview of Abraham Esau.

Esau's formal title was "Plenipotentiary of the Reichsmarschall for Nuclear Physics". With this development, Kurt Diebner would now report to Esau, and so found himself with feet in two camps: the RRC-funded KWIP in Berlin, and the Army program at Gottow. But this did not last long: Effective October 1, Werner Heisenberg was made "Director at the KWIP"; Diebner relocated to Gottow. Despite his grand title, however, Esau wielded little real political clout. In February, 1943, Albert Vögler, a favorite of Albert Speer and the President of the Kaiser Wilhelm Foundation (which funded the KWIP), informed Esau that he intended to personally apportion research between the KWIP and Gottow. The situation evolved again the next month when the War Office decided to pull out of the effort, but with the provision that Diebner's group would continue its work at Gottow with funding from the RRC.

Aside from its internal dysfunction, the German nuclear effort began to become more and more hampered by relentless Allied bombing raids. In the summer of 1943 about a third of the KWIP was relocated from Berlin to the town of Hechingen in far southern Germany. In October of that year political pressure against Esau (probably orchestrated by Speer) came to a head when Mentzel sounded out University of Munich physics professor Walther Gerlach on the possibility of his taking over the physics section of the RRC. Gerlach accepted the job, contingent on his being given absolute authority over the distribution of funds. His appointment became effective on January 1, 1944, but he permitted the two reactor groups, Diebner's and Heisenberg's, to continue operating separately until very near the end of the war; work on Heisenberg's underground pile bunker in Berlin continued. On the night of February 15, 1944, the Kaiser-Wilhelm Institute for Chemistry took a direct hit during a bombing raid, and it was decided to relocate both the Chemistry and Physics Institutes to Tailfingen, about 10 miles from Hechingen, although the pile bunker would remain in Berlin. In the late summer, Diebner's pile-research group was similarly relocated to Stadtilm, in the center of Germany.

Walther Gerlach's position was unenviable. By May 1944, Albert Vögler was making it clear that he dissatisfied with progress on the Berlin project. Gerlach decided that a safer site needed to be found, and decided on the village of Haigerloch, which lay on a river between two sheer cliffs only about 10 miles from Hechingen. A wine cellar had been carved into the rock; contracts were issued to enlarge it, but this would take several months. Finally on January 30, 1945, Gerlach ordered the Berlin group, which was planning its biggest pile yet (B-VIII; see Sec. 3.3), to evacuate. Their uranium, heavy water, and equipment were first moved to Diebner's laboratory at Stadtilm, but Heisenberg pressured Gerlach to relocate the project to Haigerloch, and the better part of a month passed before reconstruction of B-VIII got underway.

In 1943, the commander of the American Manhattan Project, General Leslie Groves, had established a vigorous program of gathering intelligence on possible German nuclear activity. This resulted in the establishment of the *Alsos* mission, a collaboration of the Manhattan Engineer District, the Army's G-2 Intelligence department, the OSRD, and the Navy. The first *Alsos* mission, commanded by Lieutenant-Colonel Boris Pash, followed the American Fifth Army as it advanced through Italy in 1943, determining that Italian scientists had done no work on nuclear explosives. In the lead-up to the planned invasion of France, *Alsos* was reconstituted, this time containing a group of scientists led by Samuel Goudsmit, a Dutch-born University of Michigan physicist. Goudsmit flew out for London on D-Day (June 6); there, he and his staff built up a target list of German scientists and industrial firms to be investigated.

Pash entered Paris on August 25, just behind the first column of French tanks to enter the city. There he found physicist Frédéric Joliot waiting for them at his laboratory, who passed on the names of Schumann, Diebner, Bothe, Bagge, and Esau as being prime movers in the German nuclear program. Another valuable clue came in the form of a catalog for the University of Strasbourg, which indicated that another high-ranking member of the German team, Carl Friedrich von Weizsäcker, was located there. Strasbourg was liberated in late November, and, upon raiding von Weizsäcker's office, *Alsos* personnel captured a trove of letters, files, and papers which listed addresses and telephone numbers for many of the uranium project's main institutes and identified Gerlach as head of the program.

In late March 1945, American troops entered Heidelberg, where Goudsmit found Walther Bothe. Bothe revealed the existence of Diebner's group in Stadtilm, that Heisenberg was in Hechingen, and that the last Berlin pile had been evacuated to Haigerloch. Stadtilm was captured on April 12, an on April 23 troops led by Pash captured Haigerloch. The German nuclear program came to an inglorious end the next day when British and American intelligence officers dismantled the pile and blew up the cave.

In contrast to the German effort, the wartime American nuclear program got off to a modest start. The first formal contact between nuclear scientists and government representatives in America occurred on March 17, 1939, when Enrico Fermi met with naval officers in Washington to explain the possibilities of using chain reactions as power sources or in bombs. Despite skepticism, the group decided to contribute $1,500 to Columbia to help advance Fermi's research on the neutron-capturing properties of graphite. This $1,500 was the first installment on an investment which would ultimately grow to nearly $ 2 billion.

The next major development occurred in October, 1939, when Alexander Sachs, an economist and advisor to President Franklin Roosevelt, delivered to Roosevelt a letter alerting government authorities to the possibilities of atomic power and bombs and the importance of securing supplies of uranium. The letter was conceived by refugee physicists Leo Szilard and Edward Teller, but signed by Albert Einstein; Szilard and Teller reasoned that, unlike themselves, Einstein would be recognized and given credibility by Roosevelt. Sachs was friend of Szilard who had some scientific training. The letter was dated August 2, and taken to the President on October 11. After hearing Sachs out, the President allegedly remarked, "Alex, what you are after is to see that the Nazis don't blow us up." Roosevelt ordered an aide to work with the Director of the National Bureau of Standards, Lyman Briggs, to put together an advisory committee. Sachs met with Briggs the next day, and they assembled the so-called Uranium Committee, which comprised Briggs and representatives from the Army and Navy. The group held its first meeting a October 21; Einstein did not attend, but Enrico Fermi, Szilard, and Teller were present. The War and Navy Departments contributed $6,000 for the purchase of four tons of graphite, paraffin, cadmium, and other supplies to support Fermi's neutron experiments at Columbia. Through the fall and winter of 1939/40, Briggs kept Roosevelt apprised via occasional reports while scientists at various institutions conducted experimental and theoretical research on fast and slow neutron reactions, uranium isotopes, and production of uranium metal. The Committee held its second meeting on April 27, by which time it had been verified that U-235 was responsible for slow-neutron fission; Fermi and Szilard were already beginning to conceive of a reactor wherein a three-dimensional lattice of blocks of uranium would be distributed within a moderator.

In the summer of 1940 the Uranium Committee underwent a significant change of venue within governmental administration. On June 27, President Roosevelt established the National Defense Research Committee (NDRC), which was charged with supporting and coordinating research conducted by civilian scientists which might have military applications. The NDRC was the brainchild of Vannevar Bush, Massachusetts Institute of Technology-trained electrical engineer who Roosevelt appointed to be its Director. During World War I Bush had worked with the National Research Council on the application of science to warfare, including development of submarines. In 1939, he became President of the Carnegie Institution of Washington, as well as Chairman of the National Advisory Committee for Aeronautics, the forerunner agency of NASA. These positions enabled him to direct research toward military applications, and gave him a conduit for providing scientific advice to government officials. Concerned over lack of cooperation between civilian scientists and the military, Bush began thinking of a federal-level agency to coordinate research, an idea he discussed with James Conant, a distinguished chemist and President of Harvard University. Bush secured a meeting with Roosevelt, and soon had his agency. Funded by and reporting directly to the President, the NDRC was remarkably free of bureaucratic interference. In addition to its involvement in the Manhattan Project, the NDRC and its successor agency, the Office of Scientific Research and Development (OSRD, July 1941), were involved with the development of radar, sonar, proximity fuses, synthetic rubber, and the Norden bomb sight. Roosevelt also directed that the Uranium Committee be absorbed into the new agency.

On July 1, Briggs summarized work to that time in a letter to Bush. Fermi's measurements of neutron capture in carbon looked promising as far as eventually obtaining a chain reaction was concerned, and it was felt that there was justification to pursue work on both methods of enriching uranium and measurements aimed at determining the feasibility of a chain reaction in natural uranium. A budget of $140,000 was authorized for further research. Between the fall of 1940 and the time of the Japanese attack at Pearl Harbor, the NDRC/OSRD had let contracts totaling about $300,000 for fission and isotope-separation research to various universities, industrial concerns, and private research organizations.

An important aspect of the American program is that it did not exist in total isolation. An equivalent group in Britain, the so-called MAUD Committee, was considering many of the same issues, and in time the formal MAUD report would have a significant impact on the American effort.

By the spring of 1941, Bush was receiving complaints about the apparently slow pace of the uranium committee's work. Bush felt that he needed some independent advice on the issue, and in April asked Frank Jewett, President of the National Academy of Sciences, to appoint a committee to review possible military aspects of fission. The work of this group, which was chaired by Nobel Laureate Arthur Compton of the University of Chicago, had far-reaching consequences.

Compton and his group met with various of the researchers involved with uranium work, and submitted their report to Jewett on May 17. While the committee felt that it would be unlikely that nuclear fission could become of military importance within less than two years, they did comment that a chain reaction could become a determining factor in warfare if it could be produced and controlled, and advocated that a strongly intensified be mounted during the following six months. Three possible military applications of uranium fission were identified: (a) production of violently radioactive materials to be used as missiles "destructive to life in virtue of their ionizing radiation," (b) as a power source for submarines and other ships, and (c) violently explosive bombs. In the latter context, it was pointed out that element 94 could potentially be produced in abundance in a chain reaction. The total cost was estimated at $350,000, mostly for pile research. Briggs soon developed an ambitious program covering pile and enrichment research budgeted at $583,000 over six months, but on June 12 the NDRC voted to allocate only $241,000 for materials despite Bush knowing that the MAUD committee in Britain was concluding that fission bombs were virtually certainly feasible.

At the June 12 meeting it was also voted to request that the NAS again review the proposed program, this time considering engineering aspects of the situation. Bush received a new report on July 15 which did not particularly address engineering issues, but rather related that experiments at the University of California at Berkeley had verified that element 94 was formed via slow-neutron capture in U-238 and that the new element underwent slow-neutron fission.

Beyond the National Academy reports, the single most important stimulus to the American fission project in the summer of 1941 came from the British MAUD report. The report was formally authorized by the Chair of the MAUD Committee, physicist George Thomson of University College London on July 15, although it was largely written by James Chadwick, discoverer of the neutron. Titled "Use of Uranium for a Bomb", the report was a remarkably accurate summary of how a fission bomb would work, down to the level of estimating the necessary critical mass. The NDRC maintained a liaison office in London, and a copy of the report was sent there and soon passed on to Bush in Washington. At about the same time, the Uranium Committee was augmented with addition of subcommittees charged with considering enrichment, power production, heavy water, and theoretical aspects of piles and bomb physics, and would henceforth be known as Section S-1 of the OSRD.

Bush briefed President Roosevelt and Vice-President Henry Wallace on October 9. Roosevelt decided that considerations of atomic policy were to be restricted to a group comprising himself, the Vice-President, Secretary of War Henry Stimson, Army Chief of Staff General George C. Marshall, and Bush and Conant, a group which would become known as the Top Policy Group. That Roosevelt had formed a group to consider nuclear policies indicated that he was beginning to clearly understand the implications of a successful uranium project. Bush described British MAUD report; Roosevelt authorized him to determine if a bomb could be made, and at what cost.

The same day, Bush requested a third National Academy Report, telling Arthur Compton that he had received a "communication from Britain" which dealt with the technical aspects of bombs. The MAUD report was available only to Bush and Conant, but Bush wanted an independent check on things to take to the President. This third report was submitted to Frank Jewett on November 17, and is a remarkable document. Over sixty pages long, various sections dealt with the conditions needed for a fission bomb, the expected effects of such bombs, estimates of how long it might take to produce them, the costs involved, and a series of recommendations for advancing the program. It was estimated that if all possible effort were spent on the program, bombs might be available in significant quantity within three or four years; the bombing of Hiroshima would occur three years and nine months to the day from the date of the report. The cost was estimated at roughly $80 to $130 million. On November 27 Bush transmitted the report to the President and the Top Policy Group; ironically, that date was about the time that a Japanese task force set sail on its mission to attack Pearl Harbor. The President authorized all of the recommendations, and Bush began augmenting S-1 with an advisory group of chemical engineers who were to begin developing plants for production plants.

The one major aspect of the project left unaddressed by the third NAS report – and directly relevant to this paper - was the possibility of developing reactors to synthesize plutonium. Bush organized a meeting of S-1 administrators for December 6 (the day before Pearl Harbor), during which Arthur Compton advocated for and was assigned responsibility for pile development. Within weeks, Enrico Fermi and his group at Columbia University would be preparing to move to Chicago. America's entry into the war the next day galvanized the uranium project. In a December 20 letter to Bush, Conant, and Briggs, Compton laid out an ambitious plan for pile work with goals of obtaining a chain reaction by October 1, 1942, having a pilot plant for the production of plutonium in operation by October 1, 1943, and to be producing useable quantities of plutonium by December 31, 1944. All of these estimates would prove reasonably durable.

On March 9, 1942, Bush sent a progress report to Roosevelt, Stimson, Marshall, and Wallace. Work was under way at full speed, and it was anticipated that by the summer, the most promising methods of enrichment and pile development would ready for pilot-plant construction. Bush urged that the whole matter should be turned over to the War Department at that time. Roosevelt responded two days later to concur with this recommendation, adding that "I think the whole thing should be pushed … with due regard to time. This is very much of the essence." Soon before the German nuclear program would be de-prioritized by Albert Speer, the American program was shifting to high gear.

Within days, Bush was informed that General Marshall had authorized Brigadier General Wilhelm Styer as the Army's contact for S-1. Styer was Chief of Staff to Lieutenant General Brehon Somervell, who commanded the Army's Services of Supply. A follow-up report to Roosevelt in June indicated that henceforth all financing would be handled by the Army's Chief of Engineers through the War Department. What would become the Manhattan Project was formally initiated when General Styer telegraphed orders to Colonel James C. Marshall of the Syracuse (New York) Engineer District to report to Washington to take command of what was being called the "DSM Project: Development of Substitute Materials. Marshall set up his first headquarters in Manhattan, and the formal name of the project later become "Manhattan Engineer District."

Some words on the wartime organization of the Army are appropriate here. By the spring of 1942 the Army was divided into three overall commands: Army Ground Forces, Army Air Forces, and the Army Service Forces (ASF), which is the department of interest here. The ASF was under the command of Lieutenant General Brehon Somervell. The Corps of Engineers (CE) was one of the operating divisions of the ASF, and the Chief of Engineers was Lieutenant General Eugene Reybold. Within the Corps of Engineers was housed the Construction Division, which was headed by Major General Thomas Robins. On March 3, 1942, Leslie Groves, then a Colonel, was appointed Deputy Chief of Construction under Robins. Groves had graduated fourth in his West Point class of November 1918, and also trained at the Army Engineer School, the Command and General Staff School, and the Army War College. By 1942, his workload was enormous: He was responsible for overseeing all Army construction within the United States as well as at off-shore bases. Camps, airfields, ordnance and chemical manufacturing plants, depots, ports, and even internment camps for Japanese-Americans all came under his purview. This experience gave him intimate knowledge of how the War Department and Washington bureaucracies functioned, and of which contractors could be depended upon to design, construct, and operate large plants and housing projects; in the spring of 1942 one of his projects was the construction of the Pentagon. While Groves later claimed in his memoirs that he was "familiar" with the atomic project in its initial stages as a part of his overall responsibilities but knew little of its details, he was in fact a very active participant from the outset, undertaking a survey of sites around the country that would have suitable power available to run the anticipated uranium enrichment plants and plutonium-producing reactors.

By September 1942, concern was growing that General (James) Marshall was not moving the project ahead fast enough, and Somervell and Styer decided to replace him with Groves. His appointment as "District Engineer" became official on the 17[th]. Within days he was attending to securing stocks of uranium ores, acquiring tracts of land at which to locate enrichment plants and piles, and contacting potential contractors. One of the latter on his list was the E. I. du Pont de Nemours and Company, which he contracted to develop and operate the plutonium-extraction plants; the firm would also eventually build the plutonium-production piles themselves.

By December 1942, Vannevar Bush was able to report to President Roosevelt that America's nuclear program was under firm and competent Army direction.

## 3. Summary of Pile Developments 1939-1946

As indicated in the Introduction, developments in France, Britain, Canada and Russia are summarized in Table I, those in the United States in Table II, and those in Germany in Table III. Detailed discussion of these Tables comprise the following sub-sections.

### 3.1 France, Britain, Canada and Russia

In Paris and its southwestern suburb of Ivry, Hans von Halban and his colleagues Frédéric Joliot-Curie, Lew Kowarski, and Francis Perrin undertook the first pile experiments essentially simultaneously with those of Herbert Anderson and Enrico Fermi at Columbia University in New York. Between March 1939 and January 1940, the Paris group constructed six piles. These involved uranium compounds, either uranyl nitrate or up to ~ 300 kg of uranium oxide plus a moderating agent (water or cubes of paraffin) contained within copper spheres of diameter 30, 50, and 90 centimeters; the spheres could be immersed in water. All pre CP-1 experiments utilized neutron sources, either radium-beryllium (Ra-Be) or radon-beryllium (Rn-Be). The French detected net neutron generation the 50 cm sphere containing a 1.6 molar solution of uranyl nitrate; this was published in the March 18, 1939 edition of *Nature*.[15] The last Paris experiment (January, 1940) involved a heterogeneous lattice-type arrangement of dry oxide plus paraffin cubes as a comparison against earlier wet-oxide homogeneous arrangements; they noted a decrease in resonant neutron capture with the heterogeneous configuration, as would be verified by Fermi. Curiously, they located their paraffin cubes *inside* the uranium. In late 1939, the French group also built, in Grenoble, a Fermi-like graphite pile, five feet square by ten feet high, to determine the neutron capture cross-section of graphite – which was apparently not low enough to indicate that it could serve as a suitable moderator.

von Halban and Kowarski relocated to Cambridge, England, with a supply of Norwegian heavy water.[16] In late 1940 they there undertook an experiment with about a ton of oxide and heavy water placed inside a 60 cm aluminum sphere which was spun at 20 revolutions per minute within a water bath to keep the oxide and heavy water well-mixed; they reported that a "potentially" divergent chain reaction was detected.[17] von Halban and Kowarski – and the 60-cm sphere - would later relocate to Canada as part of the British Mission to the Manhattan Project.

In the summer of 1939, George P. Thomson undertook experiments in London with about a ton of oxide contained within a cast-iron sphere, trying both water and paraffin as moderators; he concluded that no self-sustaining chain reaction would be possible with such moderators.

Pile development in Canada began in the summer of 1940 when George C. Laurence of that country's National Research Council along with B. Weldon Sargent of Queen's University constructed a pseudo-spherical arrangement of graphite and uranium oxide of diameter about 140 cm. Up to some 10 tons of graphite and one ton of oxide were involved; the latter was packed into seven-pound lots in paper coffee bags and distributed throughout the graphite. Laurence and Sargent achieved a neutron reproduction factor of $k \sim 0.9$; this experiment continued to about the summer of 1942. ($k = 1$ is necessary for a self-sustaining reaction. With these early piles, $k$ vales were often cited as if the size of the reactor was extrapolated to infinity.) The later heavy water moderated ZEEP (Zero Energy Experimental Pile) located at Chalk River, Ontario, achieved criticality on September 5, 1945, thereby becoming the first pile outside of the United States to do so. ZEEP comprised an aluminum cylinder 82 inches in diameter by 102 inches high which held about five tons of heavy water and 15 tons of oxide; it remained in operation until April, 1947.

The first Russian pile, F-1 ("Physics-1") went critical on Christmas Day, 1946. This device was a copy of the "305" reactor constructed at Hanford, Washington in the spring of 1944 to test fuel slugs for the plutonium production reactors (see below); as described by Rhodes (Ref. 14), the design was obtained by espionage. Operating at a steady power of 24 kW, this air-cooled spherical graphite pile was six meters in diameter and contained about 400 tons of moderator and fifty tons of pure uranium metal fuel.

### 3.2 United States

Pile experiments got underway at Columbia University in early 1939 soon after Enrico Fermi arrived there and teamed up with graduate students Herbert Anderson and H. B. Hanstein.[18] Their first effort, Columbia # 1 in Table II, involved a cylindrical tank which held $\sim 570$ kg of water and into which a neutron source could be inserted. A 13 cm diameter glass bulb could be placed in the tank; when filled with oxide, an increase in the neutron population above that present when no oxide was present was detected. My figure of $\sim 5.7$ kg of oxide in Table II is based on an estimated density of 5 gr cm$^{-3}$. Anderson, Fermi, and Hanstein estimated that each fission generated about two secondary neutrons. Their paper was published in the April 15 edition of *Physical Review*, but was dated March 16, just eight days after the submission of von Halban et al.'s first paper.[19]

The second Columbia experiment, a collaboration between Anderson, Fermi, and Leo Szilard, paralleled von Halban et al.'s second experiment. This used cylindrical cans of dimensions 5 cm diameter by 60 cm high which contained a total of $\sim 200$ kg of $U_3O_8$ immersed in $\sim 540$ liters of a manganese sulfate solution; this was presumably all contained within the same cylinder as the preceding experiment. The manganese served as a neutron detector via its propensity to capture neutrons and become radioactive. The conclusion was that a chain reaction "could be maintained in a system in which neutrons are slowed down without much absorption until they reach thermal energies …", with the proviso that "it remains an open question, however, whether this holds for a system in which hydrogen [i.e., water] is used for slowing down the neutrons."[20]

Over a year seems to have elapsed before the next Columbia pile was constructed; I designate this as Columbia A-21 after the number of the report in which it is described. This was dated September 25, 1940 and was presumably sent to the Uranium Committee of the NDRC. This experiment involved a graphite column ($\sim 3300$ kg) used for measuring the neutron slowing and capturing properties of carbon; no uranium was involved. Columbia pile A-6 (report dated January, 1941) was a re-build of A-21 with a 1.25 inch gap to permit insertion of a layer of $U_3O_8$ (~ 74 kg) to measure neutron production in uranium; it was estimated that ~ 1.7 neutrons were produced per fission.

The next Columbia pile, A-12 (report dated June 1, 1941) was a collaboration between Anderson, Fermi, and Princeton University physicists Robert R. Wilson and Edward Creutz. This involved a ~ 490 kg graphite column within which was embedded ~ 9 kg of uranium oxide held in an 8.5 cm diameter copper sphere for measurements of resonance capture by uranium. About a month later, pile A-1 was constructed; this measured 4 by 4 by 5 feet and contained about 9 kg of uranium oxide held within a copper sphere embedded within the pile. The purpose of this arrangement was to measure the absorption of thermal neutrons by uranium.

Columbia piles C-89 and C-74 (November 1941, January 1942) included no uranium; these were used to measure photoneutrons emitted by a Ra-Be neutron source and to measure the thermal-neutron absorption cross-section of boron, a common contaminant in early supplies of graphite.

In early 1942 the Fermi group began designating its creations as "exponential" piles, a reflection of the fact that the mathematical description of the density of neutrons within a pile involves exponential functions. The general purpose of these piles was to test the efficacies of various types of graphite and uranium and to try different lattice spacings and amounts of metal in uranium slugs. Fermi received graphite from three different suppliers: The National Carbon Company (AGOT and AGX brands), the Speer Carbon Company, and the U. S. Graphite Company; uranium was sourced from the Westinghouse, Mallinckrodt Chemical, and Metal Hydrides Companies. Eventually, 29 such piles would be constructed. These piles are grouped in Table II according as the reports within which they are discussed; I could find no mention of piles 4 or 12 in any of the reports I examined.

The first two such piles were constructed at Columbia in the spring of 1942 before Fermi moved to Arthur Compton's Metallurgical Laboratory at the University of Chicago. With dimensions of 8 x 8 x 11 feet (with the height of the second pile being four inches taller, according to Schwartz), these piles would have contained some 33,500 kg of graphite, assuming a density of ~ 1.6 gr cm$^{-3}$ (Fig. 2). Exponential pile # 1 is described in a report written by Anderson, Bernard Feld, Fermi, George Weil and Walter Zinn dated March 26, 1942. The column contained slots for the insertion of neutron-intensity measuring instruments, and sat atop a four foot high graphite base. Embedded within the column was a lattice of cubical iron cans eight inches on a side, each filled with about 60 pounds of $U_3O_8$. The total oxide mass would have been ~ 17,000 pounds, or 7800 kg. The reproduction constant was measured to be $k$ ~ 0.87, with the iron cans and impurities in the oxide each believed to be responsible for losses of a few percent. Exponential pile 2 is described in a very brief report dated "March, April 1942", and saw improvement of $k$ to ~ 0.918 by replacing the cans of oxide with 2160 cylindrical lumps of pressed $U_3O_8$ of average mass just under 1.8 kg each. The lumps were preheated to drive out any neutron-absorbing moisture, and the entire structure was enclosed in an iron shell to permit evacuation. This was the last Columbia pile.

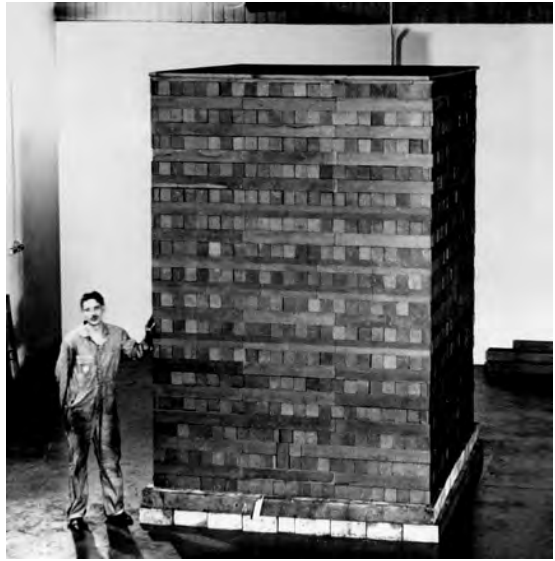

**Fig. 2.** Likely an early exponential pile. If the man is about five feet tall, this was probably an 8 by 8 by 11-foot exponential pile. *Source: https://www.atomicheritage.org/history/chicago-pile-1*

Exponential piles 3 and 5 through 9 are described in a reports dated the weeks ending June 20 and July 20, 1942. The reports do not state the sizes of the piles; presumably they were the same as their Columbia predecessors. As described in Table II, these piles used either cubes or lumps (sometimes indicted as cylindrical) of uranium oxide situated in 8-inch lattice centers. Exponential pile 7 involved the placement of two slabs of paraffin totaling 127 grams near each lump to test for loss in the reproduction factor due to the presence of hydrogenated materials; an "appreciable decrease" in the reproduction factor was detected. Exponential pile 8 was similar to number 7 except that the paraffin was replaced with beryllium; here the conclusion was that the presence of beryllium would be of no particular advantage or detriment. Exponential pile 9 featured the use of very pure Mallinckrodt $U_3O_8$, and indicated a reproduction factor greater than unity for the first time – albeit with the presence of a Ra-Be neutron source.

Exponential pile 10 used $UO_2$ in place of $U_3O_8$, which led to an increase of about 1% in $k$ to ~ 1.014. Each cylindrical slug had a mass of about 2.07 kg, although the reports do not state the total number of slugs used. Exponential pile 11 was the largest yet constructed, with a side length of 11 feet as opposed to the earlier eight feet. The height of pile 11 was not explicitly stated in the reports, but its volume is indicated as being about twice that of pile 10, which would imply a height of about 11 feet; the $k$ value of pile 11 was determined to be ~ 1.012. By the time of these results, the work had come under the auspices of the Army's Manhattan Engineer District, which took over responsibility for the project from the OSRD in August, 1942.

Paper 175 in Fermi's *Collected Works*, dated October, 1942, describes exponential piles 13-17 and 19-21 (18 is discussed below). These eight piles were built with uranium obtained from either Westinghouse (pure U metal) or Metal Hydrides ($U_3O_8$ or pure metal); all used U. S. Graphite material. Pile 13 used sintered 1.8 kg blocks of Metal Hydrides $U_3O_8$ as a comparison for other piles in this series, which involved two, four, or six "sandwich" metal-bearing layers inserted into piles which otherwise contained lumps of $U_3O_8$ with masses varying from 885 to 8848 grams each; lattice sizes were eight or twelve inches. The eight inch lattice size was determined to be about optimum; this would be used in CP-1. The best *k* value, ~ 1.035, was for pile 21, which had four sandwich layers of Westinghouse metal and 2424 gram lumps. Metal Hydrides uranium was noted as containing " … a fairly large amount of dangerous impurities … ." This series of experiments continued with piles 22-24 (November 1942). Numbers 22 and 23 used four and three sandwich layers of Westinghouse metal with lumps of mass 2727 and 3636 grams, respectively; the *k* value for both reached 1.039. These piles both used U. S. Graphite; better results were expected when National Carbon AGOT would be substituted. Pile 24 was similar to number 7 in that it was used to test the effect of incorporating paraffin in the uranium, an experiment requested by Eugene Wigner for the purpose of estimating the loss of reproduction factor to be expected in a water-cooled pile. Lumps of pressed Mallinckrodt $UO_2$ had holes bored in them to accommodate paraffin cylinders of mass 21 grams; each lump contained 1930 grams of oxide. The reproduction factor was found to be depressed by 1.9% in comparison to that of pile number 10, which otherwise had the same structure.

Piles 25 and 26 are described in a very brief report from Fermi to Leo Szilard dated the week ending October 31, 1942. The purpose of these piles was to determine what reduction in the reproduction factor would occur if large amounts of bismuth were present in the pile. The reason for this is that liquid bismuth was being considered as a coolant for the plutonium production piles. The advantage of this would be that neutron capture by the bismuth would lead to the production of polonium, which could be (and was) used in the neutron-generating triggers for the eventual nuclear weapons. Bismuth cooling was ultimately not adopted, although bismuth slugs were introduced into the Hanford piles to produce polonium in precisely this way. No details as to the dimensions or amounts of materials in the piles are given except that the amount of bismuth introduced was indicated as being about equal to the amount of uranium present. Bismuth was introduced into pile 25; pile 26 was identical except that it contained no bismuth. For the amount of bismuth projected to be used as coolant, the loss in reproduction factor was estimated to be only 0.2 or 0.3%.

Exponential piles 18 and 27-29 are described in Fermi's 1952 *American Journal of Physics* paper. All of these sat on 16 inch bases of AGX graphite, and comprised 15 layers of graphite which contained uranium alternating with 15 layers of solid graphite. Pile 18 used Speer graphite and contained pseudo-spheres of pressed Mallinckrodt $UO_2$ of average mass 2143 grams on 8.25 inch lattice centers; the total mass of oxide would have been about 3900 kg. Pile 27 was identical to pile 18 except that AGOT graphite was used throughout; this resulted in a slightly better reproduction factor (*k* = 1.039). Pile 28 was identical to pile 27 except that it was set up in a small handball court in order to make way for the construction of CP-1. The last exponential pile, number 29, was another "sandwich" design where four of the oxide-bearing layers of pile 27 were replaced with four pure metal-bearing layers with the same lattice. The uranium metal, from Metal Hydrides, was in the form of cast cylinders of diameter 2.25 inches with masses of about 2.7 kg each; the total amount of pure metal would have been about 1000 kg. The *k*-value for this pile was about 1.09.

In addition to the exponential piles, Fermi's group constructed so-called "sigma" piles for the purpose of measuring the neutron capture cross-sections of various lots of graphite. The name comes from the fact that the Greek letter sigma ($\sigma$) is used in nuclear physics to denote a reaction cross-section; the English equivalent, "*s*", serves as reminder that cross-sections express effective surface areas. A table in Ref. 8 lists 13 types of graphite, with two entries giving common cross-sections for multiple lots. All of these piles sat atop a base of Speer graphite of dimensions about 5 x 5 x 3 feet into which a neutron source could be placed in a slot which ran through the center layer of graphite. Atop the base were placed quarter-inch square graphite strips which left a gap for the insertion of neutron-absorbing cadmium. Above this would be placed 15 layers of the material to be tested, in either a cubical arrangement five feet on a side or in a rectangular parallelepiped of dimensions 168 x 157 x 157 cm (about 5.5 x 5.2 x 5.2 feet). Indium foils could be inserted into the piles to serve as neutron detectors, and all piles were sheathed in cadmium to suppress background thermal neutrons. At 1.63 gr cm$^{-3}$, a five foot cube of graphite would have a mass of ~ 5700 kg.

Construction of CP-1 began on November 16, 1942 with physicists and hired laborers working twenty-four hour days in two twelve-hour shifts under the supervision of Anderson (night shift) and Walter Zinn (day). Unlike its predecessors, CP-1 was built in the shape of an ellipsoid with an equatorial radius of 388 centimeters and a polar radius of 309 centimeters. Layers of solid graphite bricks alternated with ones within which slugs of uranium were embedded, with the slugs configured to form a cubical lattice of side length 21 centimeters as the pile was built up. This length was the average displacement over which neutrons would become thermalized after successive strikes against carbon nuclei; there would be no use in making the lattice size any larger. The bottom layer of graphite lay directly on the floor of the squash court, with the assembly supported by a wooden framework. Two special crews machined graphite and pressed uranium oxide powder into solid slugs using a purpose-designed die and a hydraulic press; the normal rate of construction was two layers per shift. In all, CP-1 incorporated 385.5 tons of graphite in the form of some 40,000 bricks averaging about 20 pounds each.

The uranium in CP-1 was in the form of pure uranium metal (just over 6 tons) and uranium oxide (about 40 tons); the slugs of pure metal were placed in the center of the pile. Holes of diameter 3.25 inches were drilled into bricks to receive the slugs, some of which were cylindrical and some pseudo-spherical; a total of 19,480 slugs were pressed.

Fermi planned for deliberate over-control of the pile. Neutron-absorbing cadmium-sheathed wooden rods could be inserted into ten horizontal slots and used to control the rate of reactivity by inserting or withdrawing them as necessary. Any one rod was sufficient to bring the reaction below criticality at any time, but Fermi insisted on redundancy. In addition, two safety "zip" rods and one automatic control rod were also incorporated into the design. During normal operation all but one of the cadmium rods would be withdrawn from the pile. If neutron detectors signaled too great a level of activity, the vertically-arranged zip rods would be automatically released, accelerated by 100-pound weights. The automatic control rod could be operated manually, but was also normally under the control of a circuit which would drive it into the pile if the level of reactivity increased above a desired level but withdraw it if the intensity fell below the desired level.

When construction was underway, all rods would be fully inserted and locked in place. Once per day, however, they would be temporarily removed and the neutron activity measured. As each layer was completed Fermi computed an effective pile radius. A plot of the square of the effective radius (a measure of the surface area of the pile, through which neutrons could escape) divided by the number of neutron counts per minute (an indirect measure of the volume of the pile) versus the number of layers was a descending curve; as the neutron flux became closer and closer to diverging, the surface-to-flux ratio would decline. By extrapolating the curve to zero, Fermi could predict the layer at which criticality would occur. By late November it was clear that the pile would become critical on the completion of its 56$^{th}$ layer; Fermi added a 57$^{th}$ of non-uranium-bearing graphite, which was laid during the night of December 1-2. CP-1 contained no neutron sources; cosmic-ray neutrons and spontaneous fissions were sufficient to initialize its operation.

The story of the startup of CP-1 is related in a number of sources.[21] Criticality was achieved at 3:36 p.m. on the afternoon of December 2. Fermi allowed the pile to operate for 28 minutes before calling for a zip rod to be inserted, at which point he estimated that the pile was operating at a power of about one-half of a Watt. Because it was uncooled and unshielded, this was the normal operating power of CP-1, although it was operated briefly at 200 Watts on December 12.

Operation of CP-1 was terminated on February 28, 1943, so that it could be dismantled and rebuilt as CP-2 at a more secluded location outside the city in the Argonne Forest Preserve. Constructed in 21 days, CP-2 achieved criticality on March 20, 1943.[22] This pile was of the more "traditional" boxlike design, and was shielded by 5 feet of concrete on all sides except the top, which was covered with about 40 inches of wood and 4 inches of lead (Fig. 3). By May, 1943, CP-2 had been operated for short times at power levels of up to 140 kW, although it was usually operated at around 1 kW. CP-2 remained in operation until May 15, 1954, and was used for studies of neutron capture cross-sections, shielding, instrumentation, and as a training facility for production operations.

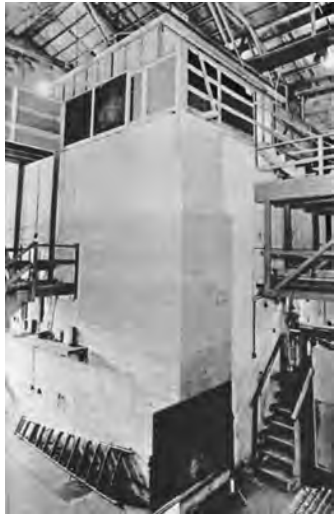

**Fig. 3.** The CP-2 pile. *Source: https://commons.wikimedia.org/wiki/File:CP-2.jpg*

The next pile to achieve criticality was the X-10 device at Oak Ridge, Tennessee; this became known as *the* graphite reactor (Fig. 4).[23] Constructed by the DuPont corporation, the purpose of this reactor was to perfect the techniques for extracting synthesized plutonium from neutron-irradiated uranium which would be used on a large scale at Hanford, and to send the extracted plutonium to Los Alamos. X-10's core comprised a 700 ton, 73-layer graphite cube about 24 feet on a side. Notches were cut into the edges of the graphite bricks so that they formed 1,248 horizontal front-to-rear fuel channels on eight-inch centers when they were laid side-by-side. Cylindrical aluminum-jacketed slugs of pure uranium metal 1.1 inches in diameter and 4.1 inches long could be fed into the channels from the front face of the pile. With the addition of concrete shielding, the outer dimensions of the pile came to some 47 feet long by 38 feet wide by 35 feet high. A full fuel load would be about 120 tons, but it was anticipated that the pile would go critical with about half that amount; the intended power level was one megawatt.

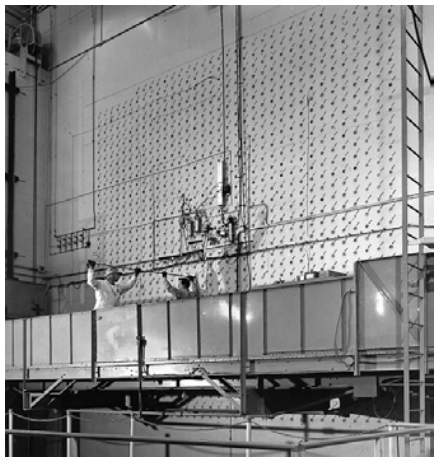

**Fig. 4.** Front face of the X-10 pile.
Source: *http://commons.wikimedia.org/wiki/File:X10_Reactor_Face.jpg*

As with CP-1, the control system for X-10 was deliberately over-designed. Three sets of control rods were incorporated: two horizontal boron-steel regulating rods, four horizontal "shim" rods, and four safety rods. The latter were boron-steel rods suspended above the pile; these could be operated manually, but were held in place with electric brakes so that they would fall into the pile in the event of a power failure. During normal operation the pile was controlled via the regulating rods, with the shim rods meant to compensate for reactivity variations too large to be handled by the regulating rods; the shim rods could effect a complete shutdown if necessary. As a backup, hoppers above the pile held small boron-steel balls which could be released into vertical channels.

Construction of X-10 began in April, 1943 and was complete by late October. Loading of fuel began on the afternoon of November 3, with Enrico Fermi inserting the first slug. X-10 went critical at 5:07 on the morning of November 4, with about 30 tons of uranium inserted. Within a week the power level had been brought to 500 kW; 800 kW was reached in December. By February 1944 the pile was producing irradiated uranium at a rate of about one-third of a ton per day; the efficiency of chemical separation of plutonium from spent fuel eventually exceeded 90 percent.

The limiting factor in X-10's operation was the capacity of its forced-air cooling system. This initially comprised two fans each capable of moving 30,000 cubic feet per minute (cfm), plus a stand-by steam-driven 5,000 cfm unit which would come on-line in the event of a power failure. By July 1944, improved fuel-slug preparation techniques and the installation of 70,000-cfm fans allowed operation at an impressive four MW. Plutonium production began in December, 1943, with a 1.5 milligrams being isolated; by mid-1944, tens of grams were being turned out per month. When production ceased in January 1945 (when the Hanford reactors were coming on-line), over 320 grams had been extracted. X-10 operated for 20 years before it was finally shut down on November 4, 1963; it is now accessible for public viewing at Oak Ridge.

The Hanford 305 reactor went critical in the spring of 1944.[24] Not well-known outside Manhattan Project history circles, this pile was a quality control tool for testing graphite, uranium, and fuel-channel aluminum tubes which would be used in the larger plutonium production piles. The core was a 16 foot graphite cube (approximate mass 300 tons) which was surrounded by foot-thick graphite reflecting blocks and a five foot thick concrete shield which could be opened for fueling and maintenance operations. The name derived from the location of the pile in the so-called "300 area" at Hanford, which was located near the town of Richland, some 25 miles from the production reactors. 305 operated until about 1973, following which it was dismantled and buried. It is still not known how plans for the 305 reactor made their way to Russia to inform the design of that country's F-1 pile.

The next reactor to go critical (May 1944) was the so called "water boiler" or LOPO (low power) installation at Los Alamos (Fig. 5). This remarkably compact device was the world's first pile which operated with enriched uranium, which by then was being produced at Oak Ridge; its mission was to provide researchers with experience in operating a chain reaction with a minimum of active material at very low power.[25,26] The core of this device was a thin-walled stainless steel sphere one foot in diameter which held a solution of uranyl sulfate ($UO_2SO_4$) surrounded by a beryllium oxide tamper and graphite neutron reflector. A cadmium rod provided for control. LOPO was operated most of the time at a power of about 50 milliwatts. Criticality was achieved on May 9, 1944, with a solution containing uranium enriched to 14.7% U-235; the amount of U-235 was 565 grams.

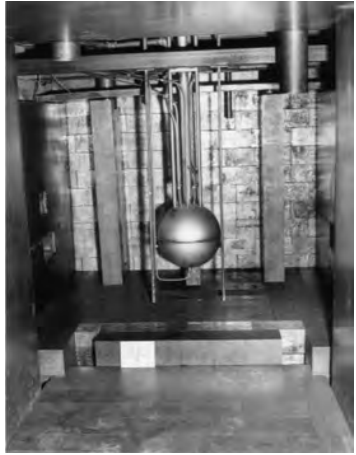

**Fig. 5.** The Los Alamos LOPO reactor.
*Source Courtesy Atomic Heritage Foundation;*
*https://www.atomicheritage.org/history/water-boiler-reactor*

Six days after LOPO achieved criticality, another milestone was reached back at the Argonne site when CP-3, the first heavy water moderated and cooled pile, achieved criticality (Fig. 6). This pile comprised an aluminum cylinder 6 feet in diameter by nine feet high which could hold about 1500 gallons (~ 6300 kg) of heavy water. Fuel was in the form of six-foot-long rods of uranium metal. The pile was surrounded by an octagonal concrete shield 13 feet high and eight fit thick, which was equipped with openings to permit inserting test materials and to access neutron beams. CP-3 operated at a power level of about 300 kW, and remained in operation for exactly 10 years until May 15, 1954, when it was shut down at the same time as CP-2. CP-3 was also known as the P-9 pile; this was the code name of the Manhattan Project's heavy water program.

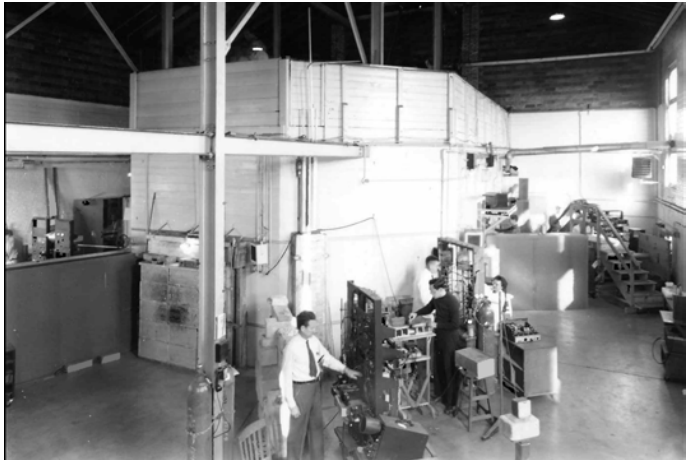

**Fig. 6.** The CP-3 pile.
*Source: https://en.wikipedia.org/wiki/Chicago_Pile-3*

At Los Alamos, LOPO's successor was HYPO, a higher-power device of the same size used to create higher neutron fluxes. HYPO was actively water-cooled, and achieved a peak power of 6 kW using an enriched uranyl nitrate solution. HYPO went critical with 808 grams of U-235 in December, 1944.

Operating at a design power of 250 MW, the plutonium production reactors at Hanford represented a leap in power output of nearly two orders of magnitude over the previous record-holder, X-10. The dramatic story of how the first Hanford reactor, B-pile, suffered crippling neutron-robbing xenon poisoning upon its startup on September 26, 1944, still stands as an example of the unexpected contingencies that can arise when developing a new technology.[27,28] The solution to this problem was to provide the reactor with more fuel, which resulted in the pile not achieving its full design power until February 1945; the companion D and F piles began their lives with full fuel loads, and commenced operation on December 11, 1944 and February 25, 1945, respectively. Shipments of plutonium to Los Alamos began in February 1945, and all thee piles operated simultaneously at 250 MW for the first time on March 28, 1945. B-pile remained in operation until February 1968.

### 3.3 Germany

The first German pile seems to have been constructed by University of Hamburg by physical chemist Paul Harteck in May-June 1940. Harteck had alerted the German War Office to nuclear developments in April 1939, and began early on to conceive of a pile where heavy water and uranium would be arranged in alternating layers. Insufficient heavy water was available, however, so he devised an ingenious substitute: Frozen carbon dioxide, known commonly as "dry ice",  which was made on an industrial scale by the I. G. Farben chemical cartel for use as a refrigerant. Harteck's experiment involved placing about 200 kg of uranium oxide in shafts drilled into a block of dry ice about six feet square by seven feet high and weighing about 15 tons. The arrangement was unrefrigerated and so lasted only a short time before the dry ice sublimated, but Harteck was able to make measurements of the diffusion length of neutrons in the dry ice and their absorption by uranium. With such a small amount of uranium he had no hope of achieving a chain reaction.

The next effort took place at the Kaiser Wilhelm Institute for Chemistry (KWIC) in Berlin in the fall of 1940, where Gottfried von Droste arranged some two tons of sodium urinate, $Na_2O(UO_3)_2 \cdot (6H_2O)$, seized from the Belgium mining firm of Union Minière de Haute Katanga (which also supplied much Manhattan Project ore) into a three-foot cube.[29] The uranate probably contained impurities; this experiment does not seem to have produced any notable results.

Activity at two other sites in Germany also got underway in late 1940. At the University of Leipzig (Werner's Heisenberg's home institution), Robert Döpel constructed a spherical pile comprising four alternating layers of water or heavy water and uranium oxide (experiment Leipzig-I, or L-I). The layers were separated by aluminum spheres, with the entire arrangement submerged in a tank of water. The aluminum spheres evidently acted as containment/support structures and were sealed with rubber; this arrangement would surely have involved some neutron loss. Results were largely inconclusive. Döpel was often assisted in experiments by his wife, Klara, a lawyer – a rare collaboration for a physics research project. It needs to be pointed out here that the numbering of L piles is somewhat confusing to reconstruct from Irving and Walker's books. That there were two four-layer piles involved in this first experiment is indicated by the latter; I have followed his designations of these as L-I (water moderator) and L-II (~ 150 kg heavy water moderator). What I designate as L-III is indicated by Irving as Heisenberg and Döpel's "second uranium pile experiment"; he clearly indicates that this piles used heavy water (his pp. 113-114) but it is not clear if this pile is distinct from Walker's L-II.

Essentially simultaneously with the Döpel's effort, Walter Bothe and Arnold Flammersfeld at the University of Heidelberg mixed together some four tons of uranium oxide and 435 kg of water in an earthenware vat, and measured the effects of resonance absorption of neutrons, concluding that an oxide/*heavy water* pile should be theoretically possible.

The next Heidelberg experiment, in January 1941, had profound consequences for the German program. Walter Bothe used a 110 centimeter diameter sphere of supposedly pure graphite to measure the diffusion length of thermal neutrons in that material. Diffusion length is a measure of how far a neutron will travel on average before it is captured; a lower diffusion length implies a higher capture cross-section, which robs a putative chain reaction of the neutrons it needs to keep propagating. Bothe determined a diffusion length of about 35 centimeters as opposed to the correct value of about 50 centimeters, and concluded that this (apparently) small value would doom the possibility of using graphite as a moderator.[30] It has been speculated that his graphite contained boron and cadmium, both voracious neutron capturers. The effect of Bothe's error was that the Germans dropped all work on graphite as a moderator, although it would be used as a "neutron reflector" in some of their piles. This forced their program to depend on supplies of heavy water from Norway for use as a moderator, a source repeatedly disrupted by Allied bombing and commando operations.[31,32]

Experiments would continue in Leipzig until mid-1942, but the focus of German pile research began to shift to Berlin in late 1940. In July of that year, planning began for a dedicated laboratory to be built on the grounds of the Kaiser-Wilhelm Institute for Biology and Virus Research, which was located next to the Kaiser Wilhelm Institute for Physics (KWIP). To deter the curious, the laboratory was designated "The Virus House". Construction was directed by Karl Wirtz, a staff member at the KWIP. The facility was ready by early October. The centerpiece of the laboratory was a two-meter-deep brick-lined circular pit which served as a receptacle for a cylindrical reactor vessel which measured 1.4 meters tall and of equal diameter. The cylinder could be lifted into and out of the pit with a crane, a more ambitious arrangement than Enrico Fermi's Columbia and Chicago piles. In December, a group directed by Wirtz and Heisenberg began assembling the first Berlin or "B" pile. The cylinder was loaded with layers of uranium oxide separated by paraffin (as a moderator), and immersed in water in the pit. A neutron source was lowered into the pile, but no chain reaction was observed; neutrons were apparently being absorbed within the pile. The experiment was repeated with 6,800 kg of uranium oxide arranged in two piles within the cylinder, but again to no avail; Heisenberg concluded that a light-water or paraffin-moderated pile would not achieve criticality.

In the meantime, work continued in Leipzig, where in the late summer/fall of 1941, Heisenberg (who shuttled between Leipzig and Berlin) and Döpel assembled some 160 kg of heavy water and 140 kg of uranium oxide in two layers within a 75 centimeter aluminum sphere, with the sphere immersed in a tank of water (pile L-III). There was no measurable neutron increase, but their confidence in eventual success began to grow. Heisenberg has been quoted as saying that "It was from September 1941 that we saw an open road ahead of us, leading to the atomic bomb."

Heisenberg's "open road" comment is cited in Irving. Irving acknowledges Heisenberg for granting him several lengthy conversations; the remark therefore probably occurred well after the fact, which makes any speculation surrounding it chancy. That the recollection refers to September 1941 is interesting as that was precisely the month of Heisenberg's infamous meeting with Niels Bohr in Copenhagen, during which Bohr inferred that the Germans were seriously pursuing construction of a nuclear weapon; might Heisenberg made a similar declaration in their conversation?[33] But in reality such a statement would have been quite a stretch. While German physicists had conceived of breeding plutonium, they were never close to doing so on an appreciable scale and apparently did little research toward the chemical issue of how it might eventually be separated from reactor fuel. In the fall of 1941 both British and American scientists were coming to the conviction that a bomb might be ready in time for use during the war, but they also realized that a very arduous and expensive journey to that goal lay ahead and that success was by no means assured. A less charitable interpretation of Heisenberg's remark is that it may represent a component of his compatriots' later rationalization that while they knew how to construct a bomb, they withheld from doing so on moral grounds; this is discussed further below. In any event, the road was not without detours: In December 1941, the Leipzig aluminum sphere was destroyed when powdered uranium, which is pyrophoric, caught fire.

A startling aspect of all German wartime piles was that they incorporated no control mechanisms.[34] This was because Heisenberg was convinced that as the temperature of a pile began to increase, it would stabilize itself at a temperature he estimated to be about 800 C. His reasoning was that as the temperature increased, thermal broadening of "resonance" energies in the uranium would lead to more neutron capture. The energy-width of resonances does increase with temperature, but this overlooks a fundamental aspect of reactor engineering: Neutron thermalization takes place *outside* the uranium in a reactor in the surrounding moderator, and typical moderating materials such as graphite, ordinary water, and heavy water have essentially no resonance-capture energies to speak of. So far as the behavior of neutron interaction with uranium is concerned, the fission cross-section for uranium-235 does decrease as the temperature rose, but at the same time the average speed of the neutrons within the reactor *increases* with temperature. These two effects largely cancel each other, leaving the time between neutron-induced fissions largely independent of the temperature. Even if his argument were valid, Heisenberg was apparently not concerned that the extreme temperature he predicted could melt the reactor (Fukushima!) or could cause a chemical explosion. In comparison to Fermi, Heisenberg's mistakenness concerning reactor characteristics is striking.

Leipzig experiment L-IV in May 1942 used pure uranium powder supplied by the firm of Degussa (German Gold and Silver Exchange). Over three-quarters of a ton of uranium and 140 kg of heavy water were encased in two bolted-together aluminum hemispheres of diameter about 80 cm, all again immersed in a water bath (Fig. 7). Results clearly indicated that more neutrons were being detected at the pile's surface than were being supplied by a Ra-Be neutron source at its center. Heisenberg and Döpel predicted that they should be able to achieve a chain-reaction if they could expand their effort to incorporate about five tons of heavy water and 10 tons of solid uranium metal. Unfortunately, catastrophe struck on June 23: L-IV was destroyed when hydrogen liberated by dissociation of some of the surrounding water exploded. Heisenberg and Döpel barely escaped without injury, and their heavy water and uranium were lost. L-IV would be the last Leipzig pile; Döpel recommended that all future piles be built with solid uranium.

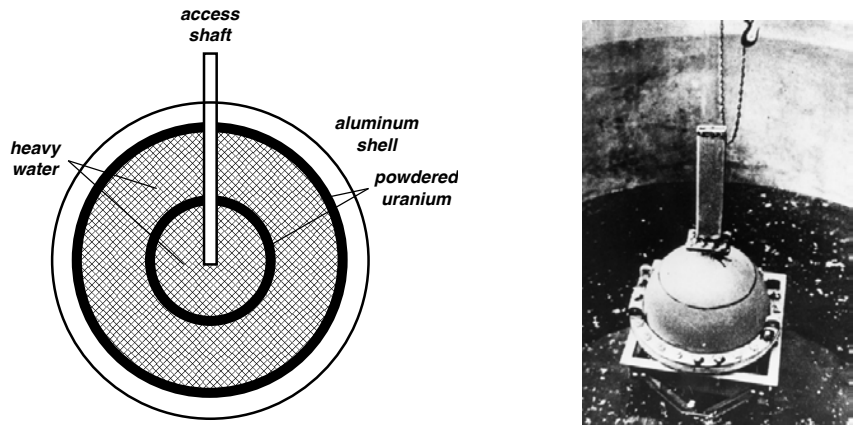

**Fig. 7.** Schematic illustration (not to scale) of the Leipzig L-IV pile. Sketch by author after Irving (Ref. 10) p. 132. Right: A German spherical pile. *Source* AIP Emilio Segrè Visual Archives, Goudsmit Collection.

The German program began to undergo significant changes in 1942, although these had been brewing since 1939. Soon after the war began, the German War Office had taken over the facilities of the KWIP. As was related in Sec. 2, the then-Director, Peter Debye, a Dutch citizen, left for a position at Cornell University in January, 1940, and was replaced by physicist Kurt Diebner, an explosives expert with the Army who was familiar with nuclear physics and also oversaw a nuclear research effort at an Army research site in the suburb of Gottow. However, KWIP staff and administrators felt that Diebner was not of the caliber of Debye, and wanted to see Heisenberg appointed instead. That did not happen at the time, although he did serve as an advisor and commuted to Berlin from Leipzig once a week. However, Heisenberg had strong allies in Carl Friedrich von Weizsäcker and Karl Wirtz, who continued to urge the governors of the Kaiser-Wilhelm Society to appoint him as director. von Weizsäcker likely carried serious influence: His father was the second-highest official in the German Foreign Ministry. They eventually prevailed; in the spring of 1942 Heisenberg became "Director at the Kaiser Wilhelm Institute for Physics" (Debye had never formally resigned); he was also appointed professor of theoretical physics at the University of Berlin.[35] Diebner departed for the Army's research site at Gottow, where he carried on with his own Army-funded pile experiments. In another key development, not long before the hydrogen explosion which destroyed L-IV, Minister of Munitions Albert Speer authorized construction of a shelter to house a large reactor on the grounds of the KWIP, a successor to the Virus House (Sect 2.3).

Sometime in the first half 1942 Diebner hit on the crucial idea of distributing uranium as chunks within a moderator, as opposed to using layers of plates or shells. He set up the first Gottow pile, G-I, that summer. Within a large cylindrical aluminum vessel, he set up a 19-layer pile wherein some 25 tons of powdered uranium oxide was poured into 6802 cubical voids cast like chambers of a honeycomb in over four tons of paraffin; the vessel was surrounded by water (Fig. 8). No increase in the neutron flux was noted, but the idea of distributing the uranium through the moderator was exactly what Enrico Fermi was doing simultaneously in Chicago.

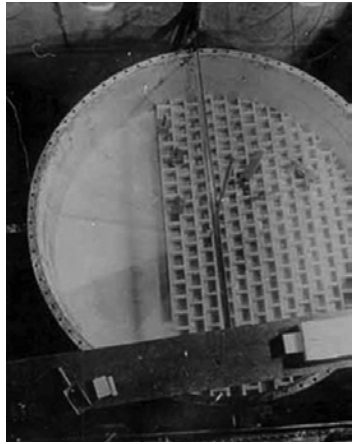

**Fig 8.** Kurt Diebner's paraffin-honeycomb G-I pile. Image courtesy Mark Walker.

At the KWIP, work on Heisenberg's preferred layered designs continued through 1942. Berlin piles B-III, IV, and V involved alternating layers of powdered uranium metal and paraffin within a sphere of about 60 cm diameter embedded in water; systems with 7, 12, and 19 layers of uranium were tried, but results were not encouraging.

In Gottow, Diebner continued to focus on lattice-type arrangements. Experiment G-II was an ingenious arrangement reminiscent of Paul Harteck's dry-ice pile of 1940. Diebner had 108 cubes of uranium 5 cm on a side made up (~ 230 kg) embedded in some 200 kg of frozen heavy water, which was itself embedded in a 75 cm sphere of paraffin wax. This was an awkward arrangement in that it made any rearrangement of the geometry time-consuming, but represented progress in that its neutron reproduction was better than any pile so far. This success directly motivated pile G-III, wherein the cubes, now 240 of them (~ 560 kg) were suspended on fine wires into a cylindrical aluminum tank which held heavy water (~ 590 kg) in liquid form. This pile exhibited 6% more neutrons leaving the pile than were supplied by the neutron source, the best result yet.

By the spring of 1944, the new laboratory in Berlin authorized by Albert Speer almost two years earlier was complete. Compared to Fermi's laboratories in America, this facility was lavishly outfitted. A circular pit housed the reactor vessel, a 124-cm wide by 124-cm tall by 3-mm thick cylinder made of low-neutron-absorbing magnesium alloy. A winch ran over the pit to raise and lower the cylinder and its lid; special ventilation, air-conditioning, and pumping equipment could siphon off radioactive emissions; the pile could be viewed though portholes; and laboratories and workshops were available for processing uranium and heavy water. Despite that fact that Diebner's cube geometry was giving superior neutron generation, the next Berlin pile, B-VI, was constructed with Heisenberg's preferred plate design. Arrangements of one-centimeter thick plates of uranium metal varying in total mass from 900 to 2100 kg were alternated with layers of heavy water (total ~ 1.5 tons) by means of spacers; the cylinder was immersed in the pit, which was filled with water. Various spacings were tried; a separation of 18 cm between the plates proved to give the best neutron increase. In contrast, in America in the spring of 1944, CP-3 with its over 6000 kg of heavy water was about to enter operation.

In late 1944, the next Berlin pile, B-VII, was constructed. A new aluminum cylinder was obtained, this one 210 cm in diameter by 210 cm tall and 5-mm thick. This would enclose the magnesium-alloy vessel of B-VI, with the space between the two filled with 10 tons of neutron-reflecting graphite. This was another plate design, incorporating some 1.25 tons of metal and the existing 1.5 tons of heavy water. No control rods were provided; Wirtz later claimed that the pile was intended to be subcritical. The neutron multiplication rate, while not yet-self-sustaining, was better than in previous experiments. Surprisingly, this did not raise any questions regarding Bothe's 1941 measurements of neutron capture in graphite.

The B-VIII pile was reconstructed in Haigerloch following its evacuation from Berlin (Fig. 9). 664 of Diebner's 5-cm uranium cubes were fixed on 78 wires suspended into 1.5 tons of heavy water; a neutron source could be inserted through a chimney in the lid. There was little instrumentation, and the only control mechanism was a block of cadmium which could be thrown into the pile if it threatened to get out of hand. The neutron flux was monitored as heavy water was pumped in, but even when the tank was full the flux had not achieved the exponential growth characteristic of a self-sustaining reaction; it was estimated that the assembly would have to be 50% larger to obtain a chain reaction, which would require yet more uranium and heavy water. But by then the game was up: On April 23, Haigerloch was captured, and the next day a group of British and American intelligence officers entered the cave and found the reactor pit. The uranium and heavy water were gone, but they dismantled the pile, seized some graphite blocks, blew up the pile's outer casing with hand grenades, and then blew up the cave. The heavy water had been hidden in gasoline cans in a country mill, and hundreds of uranium cubes were buried in a field outside the village; both were recovered.

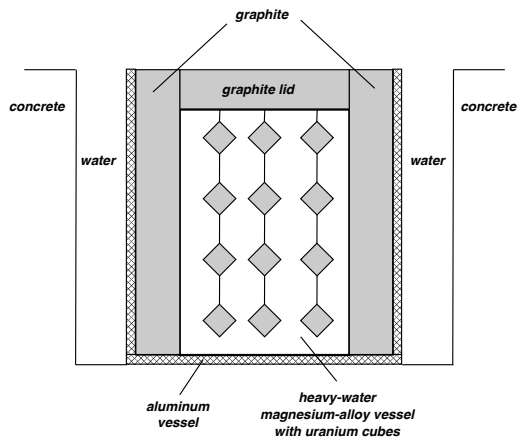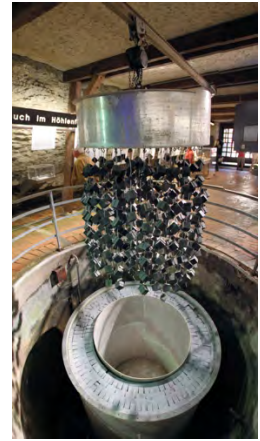

**Fig 9.** Left: Sketch (not to scale) of the B-VIII reactor, after Irving (1967) p. 319. Right: A replica of the B-VIII pile at the Atomkeller Museum in Haigerloch. *Source:* https://de.wikipedia.org/wiki/Forschungsreaktor_Haigerloch#/media/File:Haigerloch_Atomkeller-Museum_Versuchsreaktor_2013-08-18.jpg. This image is freely available for commercial use according as the terms of a Creative Commons license available at
https://creativecommons.org/licenses/by-sa/3.0/

In 2009, a group of Italian nuclear engineers published the results of an analysis of the predicted performance of the B-VIII reactor based on using software employed in the design of modern reactors.[36] This involved making various assumptions concerning parameters such as the purity of the heavy-water and graphite and the composition of the aluminum alloy in the wires used to suspend the uranium cubes. Runs with three sets of parameter choices all resulted in the reactor being subcritical, with reproduction constant values $k \sim 0.85$. The lack of criticality was apparently not due to the presence of impurities in the graphite (which was outside the inner cylindrical vessel) but was rather a geometric issue: To slow neutrons to thermal energies by having them pass through heavy-water requires a path length of about 11 cm, whereas the shortest distance between the surfaces of pairs of uranium cubes was only about 5-8 cm, depending on the direction of neutron travel. Given their emphasis on using heavy water as a moderator it is surprising that the Germans erred in this way. But by the spring of 1945 they had no time to undertake new designs.

This analysis brings up some points of comparison between the B-VIII and CP-1 piles: The former contained only about 1.5 tons of uranium, while CP-1 involved over 40 tons of material. Even with the (close) non-criticality of B-VIII, why was there such a drastic difference? Two factors are involved. First, most of the uranium in CP-1 was in the form of oxide as opposed to pure metal; oxide will have a larger elastic scattering cross-section, which translates to a higher critical mass. Second, the thermalization mean-free-path for neutrons in heavy water is about half that for neutrons in graphite: A heavy water structure can be made more compact. The CP-3 heavy water pile contained 2.8 tons of pure uranium fuel, almost twice as much as the Germans used in B-VIII.

Along a similar line, a paper published in 2015 by Klaus Mayer and his colleagues reported the results of forensic analyses of samples of metal obtained from a "Heisenberg cube" and a 1-cm thick "Wirtz plate".[37] By examining the ratios of various isotopes, they determined that both had been manufactured from ore obtained from the Joachimsthal region of the Czech Republic. Ratios of uranium isotopes indicated that the material had undergone no enrichment, and the trace amounts of plutonium present were indicative of natural origin, not as a result of any neutron irradiation. Using an analysis of the ratio of Thorium-230 to Uranium-234 (the former is the decay product of the latter), the group was able to estimate that the cube dated from the second half of 1943 (about the time of the first Gottow piles), while the plate had been produced about mid-1940. A more recent paper describes how some of the cubes made their way into private hands in America.[38]

The reaction of German scientists upon learning of their Allied counterparts' success when they were being held captive at farm Hall in England has been analyzed by a number of authors; these accounts should be required reading for anyone interested in this history.[39, 40] Particularly striking is how initial arrogant disbelief and skepticism gave way to the crushing realization that they had been far outstripped, only to then pull a form of success from the jaws of defeat by the development of a self-serving high-moral-ground rationalization that they had deliberately failed because they did not wish to see Hitler armed with atomic bombs.

4. **Concluding remarks**

By the end of 1946, a dozen piles had achieved criticality: CP-1, CP-2, and CP-3; X-10; Hanford 305, B, D, and F; LOPO; HYPO; ZEEP; and F-1. That all pile researchers would ultimately zero in on the same moderators (graphite or heavy water) and lattice or rod arrangements for the arrangement of uranium is not surprising: They were all contending with the same underlying physics, and the scientific method is such that, in time, any initial errors and misconceptions such as the unsuitability of graphite as a moderator or the self-regulatory nature of piles would be discovered and corrected.

At a straightforward level, the reasons for the failure of the German pile program in contrast to the success of the Allied one are clear: Any large, sophisticated technological project requires a thorough grasp of the scientific fundamentals, generous and consistent funding, prioritization, thousands of personnel, aggressive administration, and real support from whatever organization undertakes it. Although German scientists started from the same position of physics knowledge as Allied ones, by late 1942 they were behind in all of these factors by orders of magnitude. But there is a deeper observation to be made here as well. Just as important are the cultural, social, and political environments involved. While German scientists might have felt themselves superior to their Allied counterparts, many of the latter had studied in Germany and knew how capable German scientists and engineers were. Might fear of a German atomic bomb have kept them motivated? Allied scientists could weigh their moral convictions and choose to be involved in nuclear research or not, largely without immediate personal consequences. In the end we cannot be too judgmental of their German counterparts: By no means did they enjoy the same latitude.

**Acknowledgements**

I am grateful to David Cassidy, Miriam Hiebert, Ruth Sime and Mark Walker for comments which led to improvements in this paper. I am also grateful to Mark Walker for proving the photo of the paraffin-layered pile in Figure 8.

Table I. French, British, Canadian and Russian Piles

| Name | Location | Description/references |
|---|---|---|
| Halban et al. | Paris/Ivry | March 1939 – January 1940. Six piles; 30, 50 or 90-cm copper spheres containing uranyl nitrate solution (first pile, March 1939) or uranium oxide (up to ~ 250 – 300 kg) plus water or paraffin; could be immersed in water. Neutron generation detected, but no chain reaction. Ra-Be neutron source. Last experiment (January 1940) with heterogeneous dry oxide + paraffin cubes to compare with homogeneous wet oxide experiments; heterogeneous indicates lower resonance capture. Halban et al. (Refs. 15 - 17); Weart 116, 128, 129; Dahl 96-103. |
| Halban et al. | Grenoble | Late 1939. 5 x 5 x 10-foor graphite column for neutron-capture measurements. Dahl 102. |
| Halban et al. | Cambridge | November-December 1940. 60 cm aluminum sphere containing ~ 1 ton uranium oxide + 120 kg heavy water; immersed in water and spun at 20 rpm. Potentially divergent reaction with more material. Ra-Be neutron source. Weart 168-169; Dahl 141, 174. |
| Thomson | London | Summer 1939. ~ 1 ton uranium oxide within cast-iron sphere. Concludes no chain reaction possible with oxide and water or paraffin. Clark 35-36. |
| Laurence et al. | Ottawa | 1940-1942; George C. Laurence & B. Weldon Sargent. ~ 140 cm diameter pseudo-sphere of graphite (~ 10 tons) embedded with up to ~ 1 ton uranium oxide in 7-pound lots within coffee bags; within paraffin-lined bin. Ra-Be neutron source. Reproduction factor k ~ 0.9. Eggleston 16-26; also <https://www.nrc.gov/docs/ML0303/ML030360478.pdf> |
| ZEEP | Chalk River | Heavy water moderated an cooled. Critical September 5, 1945; shut down April 21, 1947. |
| F-1 | Russia | Critical Dec. 25, 1946. Air-cooled spherical graphite pile six meters in diameter; 400 tons graphite plus 50 tons uranium metal. Operated at 100 Watts to 1 MW; steady power 24 kW. Refs. 13 and 14 (Chapters 11, 14). |

Table II. United States Piles.  CW = Fermi Collected works, Vol. II (Ref. 7)

| Name | Location | Description/references |
|---|---|---|
| Columbia # 1 | Columbia | Spring 1939; published April 15. Cylinder 90 cm diameter by 90 cm high containing water; uranium oxide (~ 5.7 kg) within 13 cm diameter bulb. Rn-Be neutron source. Increase in neutron population detected; estimated ~ 2 neutrons/fission. Ref. 19. |
| Columbia # 2 | Columbia | Summer 1939; published August 1. 52 cylindrical cans containing total of ~ 200 kg $U_3O_8$ immersed in $MnSO_4$ solution. Ra-Be neutron source. Speculation that chain reaction possible. Ref. 20. |
| A-21 | Columbia | CW paper 136, Sept. 25, 1940. Graphite column 3 x 3 x 8 feet; no uranium. To test slowing and capture of neutrons by carbon. Rn-Be neutron source. |
| A-6 | Columbia | CW paper 138, Jan. 17, 1941. Re-built A-21 with gap to insert $U_3O_8$. Estimated ~ 1.7 neutrons produced per each captured. Rn-Be neutron source. |
| A-12 | Princeton | CW paper 139, June 1, 1941. 61 cm x 61 cm x 81 cm graphite column; ~ 9 kg uranium oxide within 8.5 cm radius copper sphere within pile to test resonance capture of neutrons by uranium. Neutrons produced by bombardment of Be with 8 MeV protons in Princeton cyclotron. |
| A-1 | Columbia | CW paper 142, July 3, 1941. 4 x 4 x 5 foot graphite pile. Copper spheres containing uranium oxide (9.3 kg $U_3O_8$ in 8.5 cm radius sphere) or powdered metal (5.5 kg in 5.65 cm radius sphere) placed within pile to measure absorption of thermal neutrons by uranium. Rn-Be neutron source. |
| AP-89, C74 | Columbia | CW papers 147 and 148; Nov. 1941 and Jan. 1942. 4 x 4 x 7 and 4 x 4 x 6 foot graphite columns; no uranium. To measure photoneutrons emitted by beryllium irradiated with gamma-rays from radium and the absorption cross-section of boron for thermal neutrons. |

Table II (Continued).

___

| Name | Location | Description/references |
|---|---|---|
| Exponential 1, 2 | Columbia | CW papers 150, 151; January and March 1942. 8 x 8 x 11 foot and 8 x 8 x 11' 4'' graphite columns. #1 comprised ~ 7800 kg uranium oxide in 288 iron cans to measure neutron production in oxide/graphite lattice; column atop 4 foot graphite base. Reproduction constant $k \sim 0.87$. # 2 included ~ 3900 kg pressed uranium oxide cylinders ~ 1.8 kg each; 16-inch graphite base; reproduction constant $k \sim 0.918$. #2 was the last Columbia pile. |
| Exponential 3, 5-9 | Chicago | CW papers 164, 165, weeks ending June 20 & July 25, 1942. Size of piles not stated in reports. Various configurations involving 3 inch oxide cubes in 8 inch lattice, 18 kg $U_3O_8$ lumps, oxide cubes with paraffin nearby, and oxide cubes with beryllium nearby. AGX and US Graphite materials. # 9 used very pure Mallinckrodt $U_3O_8$ and indicated $k \sim 1.004$ if extrapolated to infinite size. |
| Exponential 10, 11 | Chicago | CW papers 166-168, month ending Aug. 15, 1942, week ending Aug. 29, 1942, and month ending Sept. 15, 1942. Pile 10 was as pile 9 but using $UO_2$ instead of $U_3O_8$. Pile 11 was the largest yet constructed, with dimensions of 11 x 11 x 11 feet (height inferred), which would have involved ~ 61,000 kg of graphite. |
| Exponential 13-21 | Chicago | Not including exponential pile 18 (see below). CW Paper 175, month ending October 15, 1942. Various numbers of layers of pure uranium metal inside piles with lumps of $U_3O_8$ of various masses and lattice sizes (mostly eight inches); see text for details. |
| Exponential 22-24 | Chicago | CW paper 178; month ending Nov. 15, 1942. Piles 22 and 23 were continuations of 19-21 above, using three or four layers of Westinghouse metal lumps of $U_3O_8$ in eight inch lattices; U. S. Graphite. Pile 24 was the same structure as # 10 but used to test the effect of paraffin inside Mallinckrodt $UO_2$ lumps. |

Table II (Continued).

| Name | Location | Description/references |
|---|---|---|
| Exponential 25, 26 | Chicago | CW paper 177, Oct. 31, 1942. To test effect of presence of bismuth on reproduction factor. |
| Exponential 18, 27-29 | Chicago | Ref. 8; CW paper 178, month ending Nov. 15, 1942. Oxide (18, 27, 28) or oxide plus pure uranium slugs in a graphite columns (Speer or AGOT) of dimensions 99 by 99 by 123.75 inches sitting on bases of AGX graphite. Mallinckrodt $UO_2$; Metal Hydrides pure uranium. Lattice spacing 21 cm. |
| Sigma piles | Chicago | See Ref. 8. Probably fall 1942. ~ 5 x 5 x 5 foot piles for measuring neutron capture cross-sections of various lots of graphite. |
| CP-1 | Chicago | Critical Dec. 2, 1942. 385.5 tons of graphite, 6.2 tons pure uranium metal, 40.3 tons uranium oxide. Operated until end of February, 1943. Maximum power 200 Watts. |
| CP-2 | Argonne | Critical March 20, 1943. See CW paper 187 and Ref. 22. 30 x 32 x 21 feet; 472 tons graphite plus 52 tons uranium. Operated until May 1954. |
| X-10 | Oak Ridge | Critical Nov. 4, 1943. Core 24 feet square by 24 feet, four inches high. 700 tons graphite; fuel load up to 120 tons pure uranium. Operated at up to 4 MW. Forced-air cooling. Shut down Nov. 4, 1963. |
| Hanford 305 | Hanford | Critical appx. March 1944. 16 foot graphite cube with 292 fuel channels; 8.5 inch lattice. For testing graphite and fuel for Hanford plutonium production reactors. Usually operated at < 50 W. See Refs. 13 and 22. |
| LOPO | Los Alamos | Critical May 9, 1944. First pile with enriched uranium (14.7% U-235); uranium sulfate solution in steel sphere of diameter one foot. Power level ~ 50 mW. Ref. 25, pp. 182, 199-203; also Ref. 26. |

Table II (Continued).

| Name | Location | Description/references |
|---|---|---|
| CP-3 | Argonne | Critical May 15, 1944. First heavy water moderated and cooled reactor. Aluminum cylinder 6 feet diameter by nine fete high; 120 rods of uranium metal six feet long by 1.1 inches diameter (total ~ 2500 kg). Normal operating power 300 kW. See Ref. 22 and Weart (Ref. 2) p. 189. |
| HYPO | Los Alamos | Successor to LOPO. Critical December 1944. 15.5% U-235 uranyl nitrate solution in steel sphere of diameter one foot. Peak power 6 kW; water cooled. Ref. 26. |
| Hanford B, D, F | Hanford | 250 MW plutonium production piles. 250 tons of uranium metal slugs passing through 38 x 38 x 28 foot graphite core in 2004 aluminum fuel channels. Water-cooled. Each pile was theoretically capable of producing about 180 grams of plutonium per day when operating at full power. |

Table III. German piles.

| Name | Location | Description/references |
|---|---|---|
| Dry-ice pile | Hamburg | May-June 1940. Paul Harteck, University of Hamburg. ~ 200 tons uranium oxide in ~ 15 ton block of dry ice. Pile ~ 6 feet square by seven feet high. No neutron multiplication detected. Irving pp. 66-69; Walker pp. 25-26; Dahl p. 142. |
| Berlin | Berlin | Gottfried von Droste, fall 1940. ~ 2 tons sodium uranate in ~ 2000 packages arranged into ~ 3 foot cube. Irving pp. 86-87. |
| Leipzig L-I, L-II | Leipzig | Late 1940/early 1941. Robert & Klara Döpel. Concentric spheres of water or heavy water moderator alternating with uranium oxide. Irving p. 89 indicates that paraffin was used as a moderator; Walker pp. 40-41 indicates both water and heavy water were used. See illustration Walker p. 40. For L-II, Dahl p. 189 gives 164 kg heavy water + 142 kg uranium oxide; outer diameter 75 cm. |
| Heidelberg | Heidelberg | Late 1940? Walter Bothe & Arnold Flammersfeld; four tons uranium oxide plus 435 kg water. Conclude than an oxide/heavy water pile should be possible. Irving p. 89 describes the oxide as "black", which is usually used to designate $U_3O_8$. |
| Heidelberg | Heidelberg | January 1941; Walter Bothe. 100 cm diameter graphite sphere. Concludes that graphite will not be feasible as a moderator, forcing the German program to rely on heavy water as a moderator. Bothe's graphite must have contained impurities. Irving pp. 92-94. |
| B I, B-II | Berlin | Late 1940. "Virus House" piles. Layers of uranium oxide (~ 6800 kg total) and paraffin within a cylindrical vessel immersed in water. Irving pp. 88-89. See illustration Walker pp. 39; he gives the surrounding water vessel as two meters diameter by the same height. No chain reaction. Also Dahl p. 188. |
| L-III | Leipzig | Late summer/fall 1941. Uranium oxide + heavy water in two layers within aluminum sphere. Irving pp. 113-114; 130; Dahl p. 189. Dahl gives 164 kg heavy water + 108 kg metal. Destroyed by fire late 1941. |
| L-IV | Leipzig | May/June 1942. ~ 0.75 tons pure uranium plus 140 kg heavy water in aluminum sphere. Net neutron generation detected. Destroyed by hydrogen explosion June 23, 1942. Irving pp. 130-131; 137-139; Dahl p. 190-191. |

Table III (Continued).

| Name | Location | Description/references |
|---|---|---|
| G-I | Gottow | Kurt Diebner; summer 1942. 25 tons powdered uranium oxide in honeycombed layers of paraffin (4 tons). Irving pp. 167-168; Walker pp. 95-96; Dahl p. 210. |
| B-III, IV, V | Berlin | 1942. Alternating layers of powdered uranium metal and paraffin within ~ 60 cm diameter sphere. Walker pp. 53-54; Irving p. 168; Dahl p.189. |
| G-II | Gottow | April 1943. 5 cm cubes of uranium embedded in ~ 600 liters frozen heavy water. Walker pp. 99-100; Irving pp. 201, 221; Dahl p. 210. |
| G-III | Gottow | Late 1943. 5 cm cubes of uranium suspended in heavy water. Walker p. 104; Irving pp. 223, 374; Dahl p. 221. |
| B-VI | Berlin | Spring 1944. Alternating layers of uranium metal plates and heavy water within magnesium-alloy cylinder. Irving pp. 274-276; Walker p. 85. Dahl p. 223 indicates four different spacing geometries, B-VIa through B-VId, with between 0.9 and 2.1 tons of uranium metal. |
| B-VII | Berlin | Late 1944. B-VI cylinder within larger aluminum cylinder, separated by graphite. Uranium metal plates plus heavy water. Irving pp. 312-314; Walker p. 85; Dahl p. 223. |
| B-VIII | Haigerloch | February – April 1945. 664 uranium cubes suspended in heavy water, surrounded by graphite. Destroyed by Allied forces April 1945. |

# This is a footnote reference list